\documentclass[12pt,notitlepage,a4paper]{article}
\usepackage{color,graphicx}
\usepackage{amssymb,tcolorbox}
\usepackage{delarray,amsmath,bbm,tikz}
\usepackage[american]{babel}
\usepackage{enumitem}
\usepackage[utf8]{inputenc}
\usetikzlibrary{arrows, arrows.meta}
\usepackage{caption,float}
\usepackage{subcaption}
\usepackage{adjustbox}
\usepackage{authblk}

\newcommand{\be}{\begin{equation}}
\newcommand{\ee}{\end{equation}}
\newcommand{\beq}{\begin{equation}}
\newcommand{\eeq}{\end{equation}}
\newcommand{\beqa}{\begin{eqnarray}}
\newcommand{\eeqa}{\end{eqnarray}}

\newcommand{\bear}{\begin{eqnarray}}
\newcommand{\eear}{\end{eqnarray}}

\numberwithin{equation}{section}

\newfont{\namefont}{cmr10}
\newfont{\addfont}{cmti7 scaled 1440}
\newfont{\boldmathfont}{cmbx10}
\newfont{\headfontb}{cmbx10 scaled 1728}

\newtheorem{prop}{Claims:}

\title{Incomplete RG: Hawking-Page transition, C-theorem and relevant scalar deformations of global AdS }
\author[*]{Pallab Basu\thanks{pallab.basu@wits.ac.za}} 
\author[**]{Pranav Kumar\thanks{pranav.kumar@gmail.com}} 
\author[*]{Chuene Mosomane\thanks{956785@students.wits.ac.za}}
\affil[*]{School of Physics, University of Witwatersrand, Johannesburg, South Africa}
\affil[**]{Haber, Pune, India}

\begin{document}

	\maketitle
		\abstract{\small We discuss relevant scalar deformations of a holographic theory with a compact boundary. An example of such a theory would be the global AdS$_4$ with its spatially compact boundary $S^2$.  To introduce a relevant deformation, we choose to turn on a time-independent and spatially homogeneous non-normalizable scalar operator with  $m^2 = -2$. The finite size of a compact boundary cuts down the RG flow at a finite length scale leading to an incomplete RG flow to IR.  We discuss a version of {\it incomplete} C-theorem and an {\it incomplete} attractor like mechanism. We discuss the implication of our results for entanglement entropy and geometric quantities like scalar curvature, volume and mass scale of fundamental excitation of the how these quantities increase or decrease (often monotonically) with the strength of the deformation. Thermal physics of a holographic theory defined on a compact boundary is more interesting than its non-compact counterpart. It is well known that with a compact boundary, there is a possibility of a first order Hawking-Page transition dual to a de-confinement phase transition. From a gravity perspective, a relevant deformation dumps negative energy inside the bulk, increasing the effective cosmological constant ($\Lambda$) of the AdS. Dumping more negative energy in the bulk would make the HP transition harder and the corresponding HP transition temperature would increase. However, we have found the size of the BH at the transition temperature decreases.} 	
 
\section{Introduction}

 Holographic C-theorem \cite{Myers:2010tj} and its relatives state that C-function  decreases in a holographic RG flow \cite{Akhmedov_1998}. That is, a deformation of a holographic theory by a relevant operator in UV initiates a flow to an IR theory with a lesser value of C-function (often the central charge) than the UV theory. This change in C-function is monotonic along the RG flow.  We have,
 \begin{align}
 C_{UV}>C_{IR}
 \end{align} 
 Let's assume both UV and IR theory are described by conformal fixed points and have dual descriptions in terms of a gravity/string theory in an AdS space \cite{Aharony:1999ti}.  At the level of the GR solution, a flow is an interpolation from a UV AdS to IR AdS. For AdS$^5$, $C \propto L^3$ where $L$ is the AdS radius \cite{Myers:2010tj}. Hence it is guaranteed that $L_{UV} > L_{IR}$, or $\Lambda_{IR}>\Lambda_{UV}$, where $\Lambda$ is the cosmological constant. In most symmetric gravity setups, holographic RG flow is a radial flow which has information about how a given AdS solution changes when we venture deep inside the bulk from near the boundary.
 
 We expound a little on the finite temperature physics. The thermal physics of a flat CFT is straight forward. There is no length scale and turning on any temperature creates a dual black hole right up to zero temeprature. An example of such a BH would be flat AdS black hole defined in the Poincare patch AdS. Deforming the relevant CFT changes its low temperature physics in an interesting way. Physics of the high temperature phases is governed by the UV fixed point and its low temperature physics is determined by the IR fixed point. However, there is no phase transition between low and high temperature phases. In the gravity side, at low temperature the black hole looks like a black hole in IR AdS. This black smoothly transitions to a BH in UV AdS as the temperature is increased \cite{G_rsoy_2018}.
 
  In most discussions of holographic RG flow and C-theorems the holographic boundary is assumed to be flat. An interesting question to ask is the consequence of a compact boundary, which provides a cutoff at the size of the boundary and the boundary theory does not have room to flow to IR. In the gravity side, with or without the deformation, one or more compact boundary dimension/s shrinks to zero inside the bulk, and the theory never flows to IR. For example, gloabl AdS$_{d+1}$ with a compact spatial boundary $S^3$ has a distinct center in it's bulk. The AdS$_{d+1}$  metric near the origin is like that of the flat space and is given by,
  \begin{align}
  ds^2 \approx -dt^2+dr^2+r^2 d\Omega_{d-1}^2,
  \end{align}
  displaying a regular shrinking of $d-1$-sphere near the origin of the polar co-ordinate. 
  
  \paragraph{}
What happens if such a theory defined on a compact boundary is perturbed with a relevant operator in UV? From the gauge theory intuition, clearly such a theory can not flow to IR, as there is no room. This resembles an incomplete RG flow and is dual to an incomplete attractor mechanism in the bulk, where bulk fields do not fully flow to the minimum potential configuration deep inside IR.
   \begin{figure}
   	\begin{center}
   		 \includegraphics[scale=0.5]{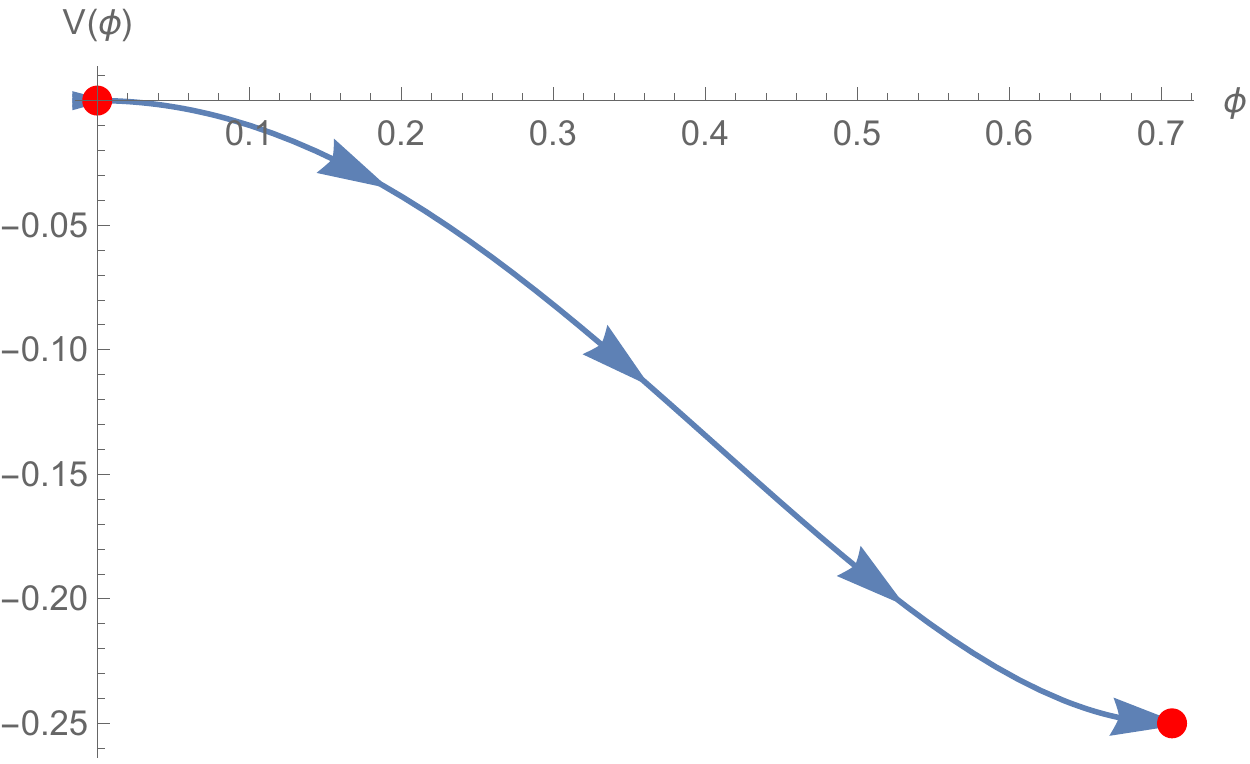}
   		\includegraphics[scale=0.5]{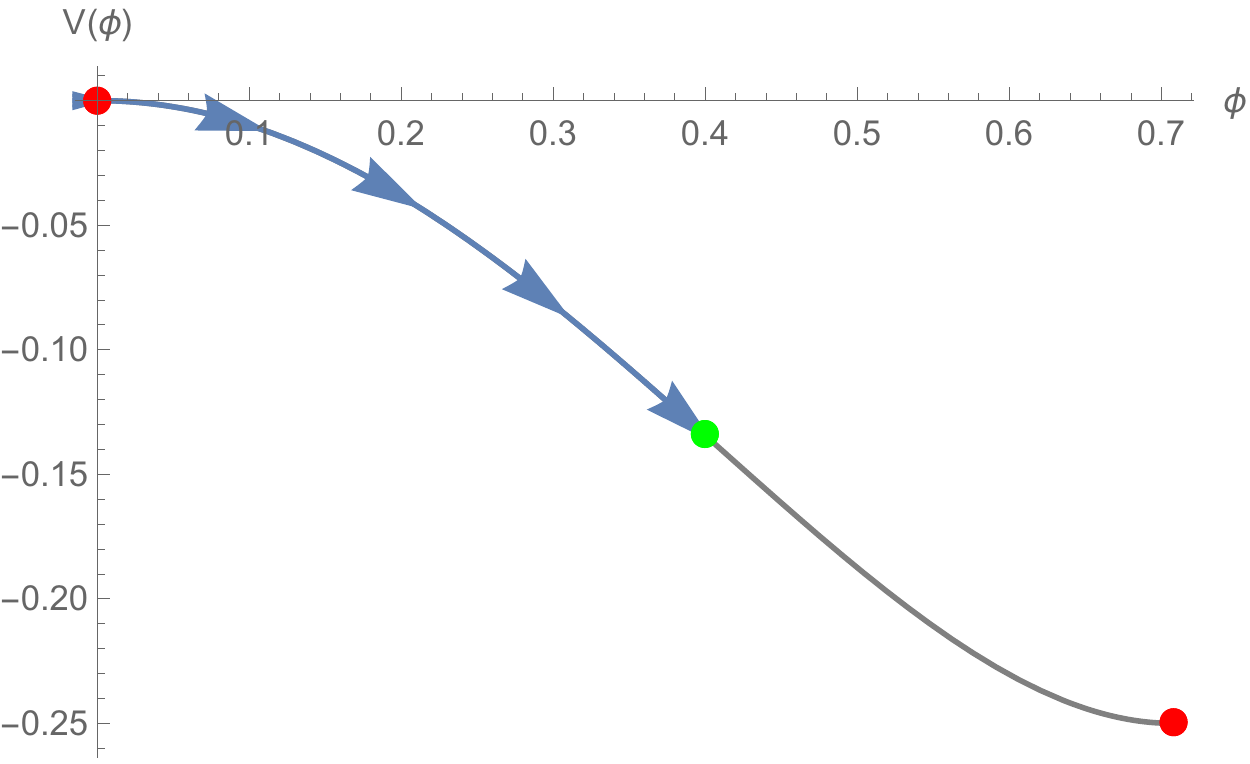}
   	\end{center}
   \caption{Incomplete RG: Schematic diagram of a UV to IR radial flow with a scalar operator with bulk potential $V(\phi)$. The complete RG flow in the flat boundary case (left) vs incomplete RG flow with a compact boundary (right). The flow in the later case terminates at the center of the space (the green point).}
   	\end{figure}
    At the level of Einstein's equations, one intuition we may develop from the study of holographic C-theorems for the flat case is that a relevant perturbation would dump negative energy inside the bulk. In the compact boundary case, the negative energy can't escape and would be contained near the center of the bulk. This would lead to an anti-gravity effect and may have the following consequences:
   
\begin{prop} \label{item:main}
	\hphantom{1}
	 \newline
	\begin{enumerate}
	\item{\bf HP transition temperature may increase. Entropy difference at HP transition temperature may decrease.} 
	
	Thermal physics of a holographic theory defined with a compact boundary is more interesting than its non-compact counterpart. With a compact boundary we have a possibility of a Hawking-Page like de-confinement transition (HP)\cite{Witten:1998zw,hp}. This is a consequence of having a mass-gap in the boundary due to compactness. These kind of transition is often first order. In our case, dumping negative energy in the bulk makes formation of BH harder. Hence under a relevant deformation of the boundary theory, HP transition temperature may increase. Whereas the radius of the BH is expected to decrease at HP transition temperature. To note: entropy difference of two phases at the transition temperature is the area of the BH at that temperature.  
	
	 \item A relevant deformation may dump more negative energy inside the space time leading to decrease in effective volume  and increase in cosmological constant ($\Lambda$) of the bulk . 
	 \item Mass scale of fundamental excitations of  AdS is determined by $L$ or $\Lambda$ as $m^2 \propto 1/L^2 \propto \Lambda$. Hence any fundamental frequency associated with the AdS may increase if $\Lambda$ is increased. 
	 \item {\bf Entanglement entropy} If we look at entanglement (EE) of a boundary region with an area $A$, then $EE = c_{UV} A/a^2+\delta EE$, where $a$ is an UV cutoff. $c_{UV}$ is closely related to a central charge and at finite $J$, $\delta EE(J)<\delta EE(0)$. As we take bigger and bigger area, $\delta EE$ would increase in magnitude and then saturate for the largest area. As the space is compact the the area A may not be arbitrarily large. Largest value of $A_{max}$ is  half big as the boundary area. For a boundary $S^n$, the largest area is given by the half-sphere. 
	 \item We also expect: as we turn on a source J, $\delta EE$ would decrease monotonically with J. This behavior should be observed in easily calculable EE of  $A_{max}$.
	  	\end{enumerate}    
\end{prop}
 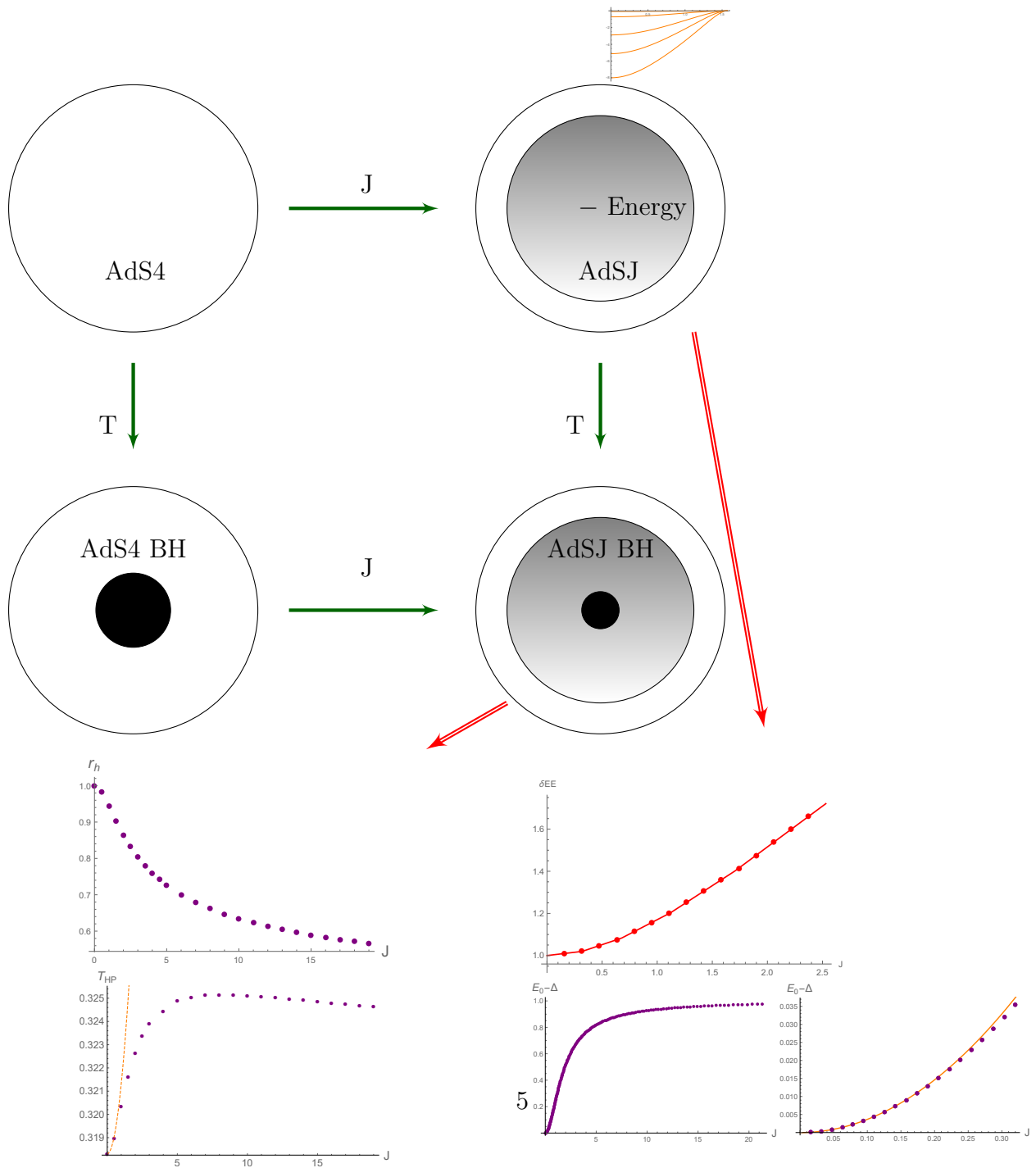
\begin{figure}\label{fig:papermap}
 	\caption{A schematic map of the paper showing the effects of turning on $J$. From left, reduction in the size of BH at HP transition. incerase in HP transtion temp, increase in $\delta_{EE}$, increase in the frequency of fundamental modes.}

 	\begin{subfigure}{10cm}
 		\hspace*{9.5cm}
 		\includegraphics[width=2cm]{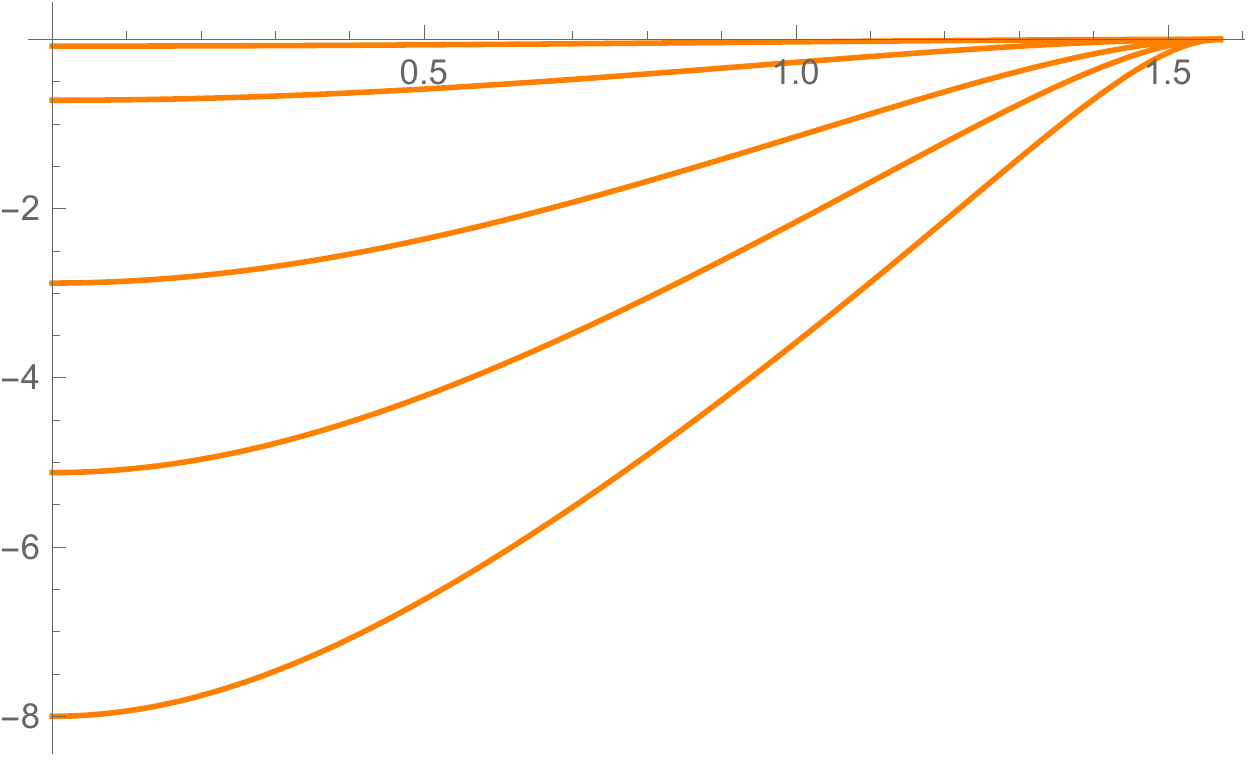}
 		 \begin{tikzpicture}
 		\draw (2.5,2) circle (2cm);
 		\draw [ultra thick, green!40!black, -latex'] (5,2.) -- +(0:2.4);
 			\draw [ultra thick, green!40!black, -latex'] (5,-4.5) -- +(0:2.4);
 		\draw (10,2) circle (2cm);
 		\draw (2.5,-4.5) circle (2cm);
 		\draw[fill] (2.5,-4.5) circle (0.6cm);
 		\draw [ultra thick, green!40!black, -latex'] (2.5,-.5) -- +(-90:1.4);
 		\draw [ultra thick, green!40!black, -latex'] (10,-.5) -- +(-90:1.4);
 		\draw (10.0,-4.5) circle (2cm);
 		\draw[shade] (10,2) circle (1.5cm);
 		\draw[shade] (10,-4.5) circle (1.5cm);
 		\draw[fill] (10,-4.5) circle (0.3cm);
 		\filldraw[black] (1.8,-1.5) circle (0pt) node[anchor=west] {T};
 		\filldraw[black] (9.3,-1.5) circle (0pt) node[anchor=west] {T};
 		\filldraw[black] (1.9,1.0) circle (0pt) node[anchor=west] {AdS4};
 		\filldraw[black] (1.5,-3.5) circle (0pt) node[anchor=west] {AdS4 BH};
 		\filldraw[black] (6,2.4) circle (0pt) node[anchor=west] {J};
 		\filldraw[black] (6,-3.8) circle (0pt) node[anchor=west] {J};
 		\filldraw[black] (9.5,1.0) circle (0pt) node[anchor=west] {AdSJ};
 		\filldraw[black] (9.0,-3.5) circle (0pt) node[anchor=west] {AdSJ BH};
 		\filldraw[black] (9.5,2) circle (0pt) node[anchor=west] {$-$ Energy};
 		\draw [double,thick, red!40!red, -latex'] (8.5,-6.) -- +(-150:1.5);
 		\draw [double,thick, red!40!red, -latex'] (11.5,0.) -- +(-80:6.5);
 		\end{tikzpicture}
 	\end{subfigure}
 \newline
 \vspace*{-5cm}
\hspace*{1cm}\begin{subfigure}{5cm}
	\includegraphics[width=5cm]{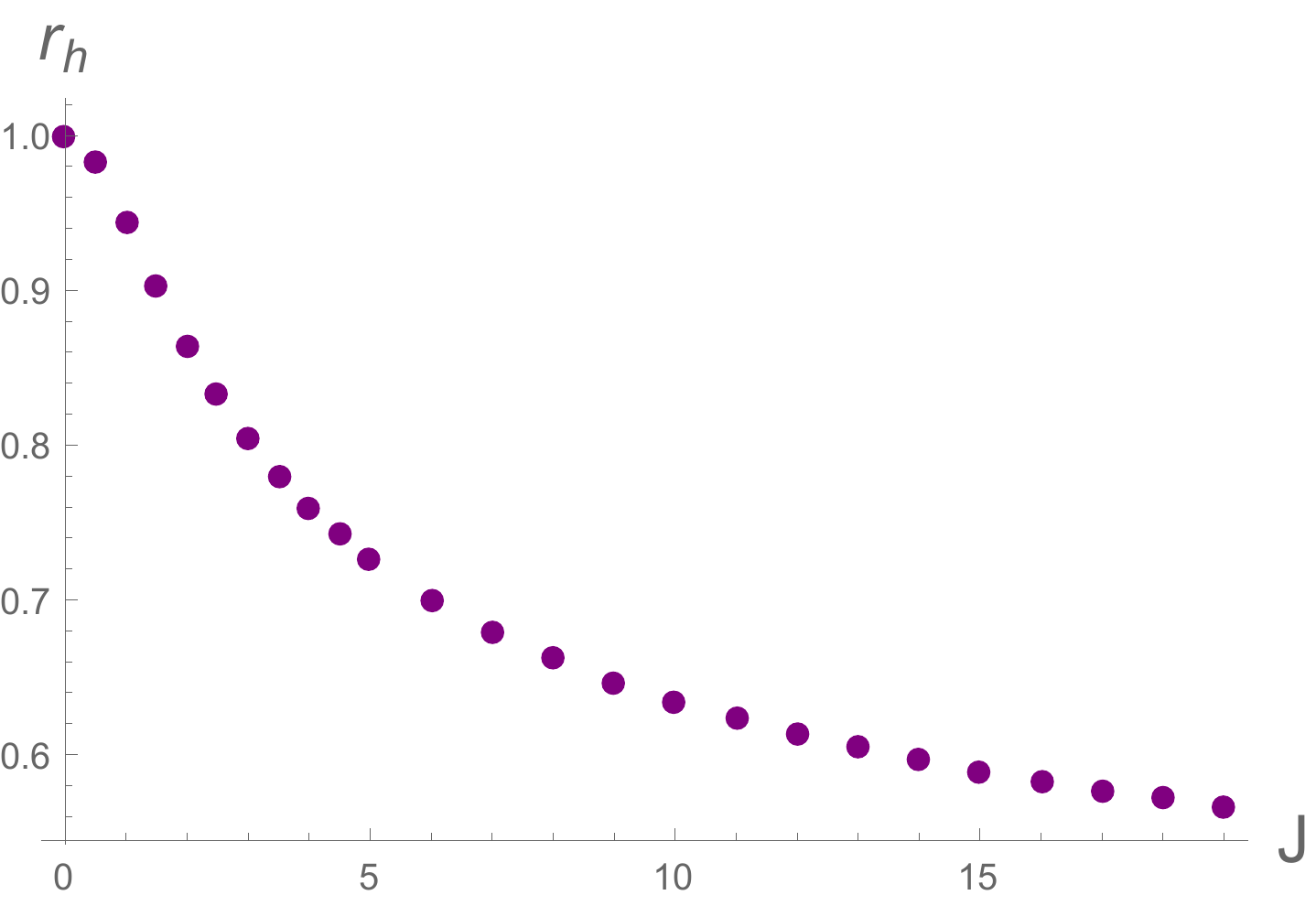}
	\newline
 	\includegraphics[width=5cm]{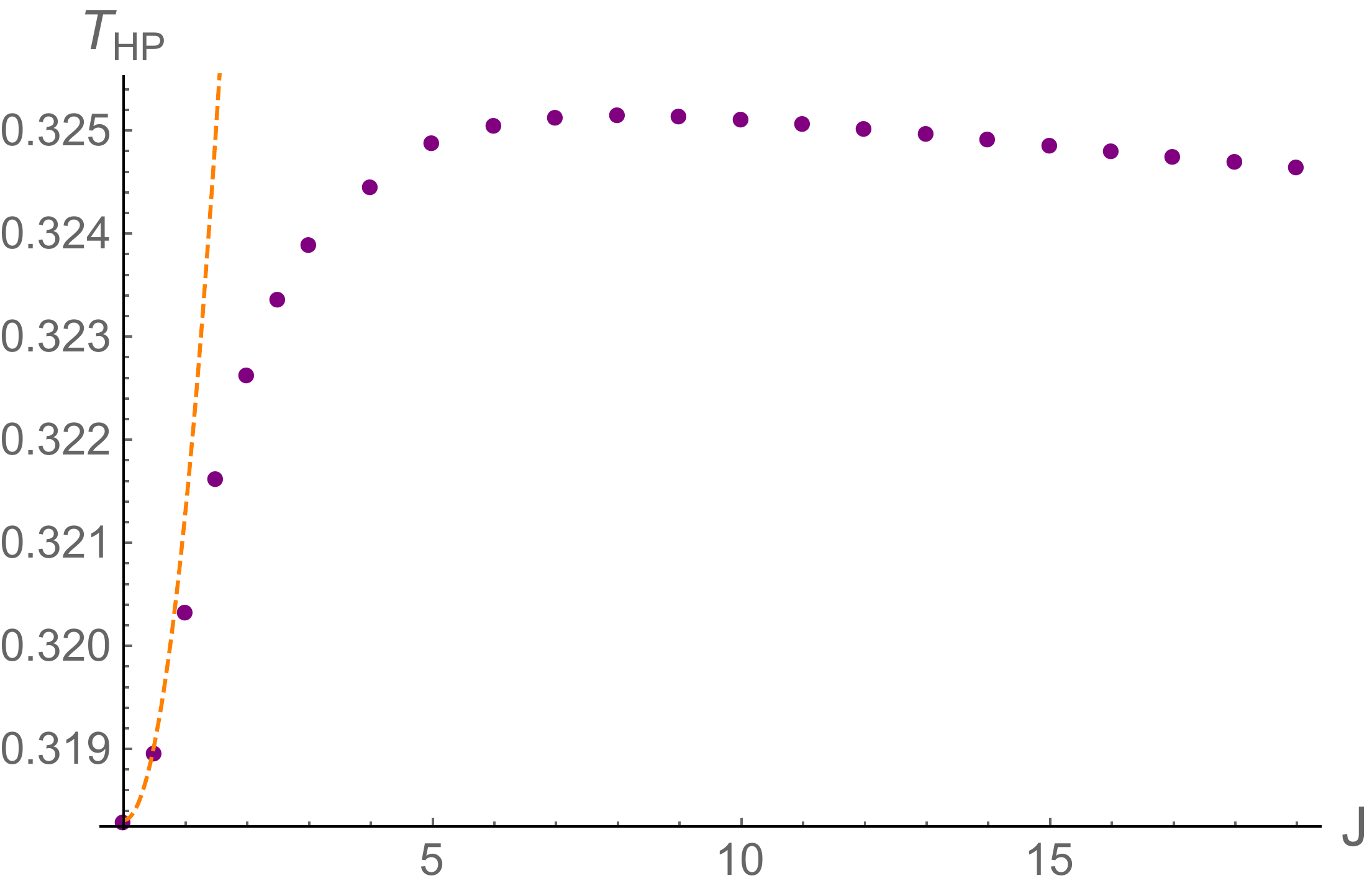}
 \end{subfigure}
\hspace{2cm}
\begin{subfigure}{4cm}
	\includegraphics[width=5cm]{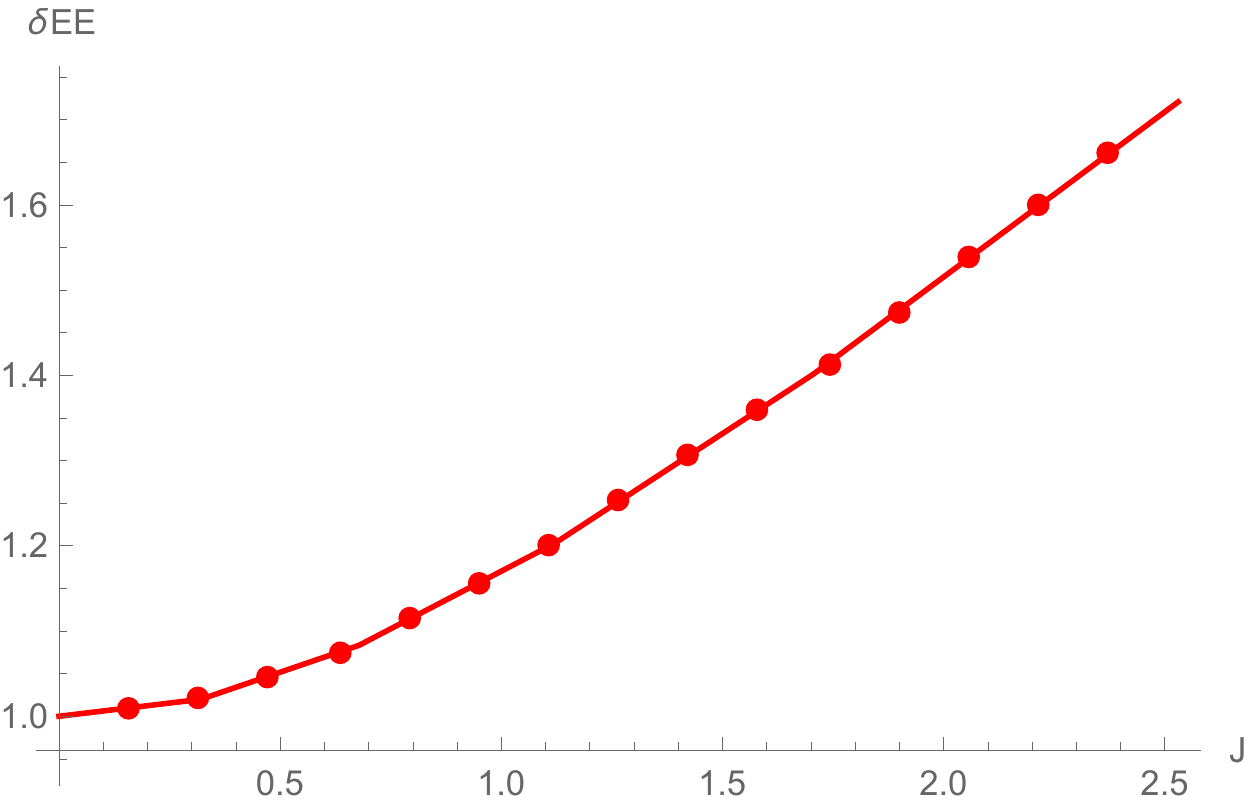}
	\newline
	\includegraphics[width=8cm]{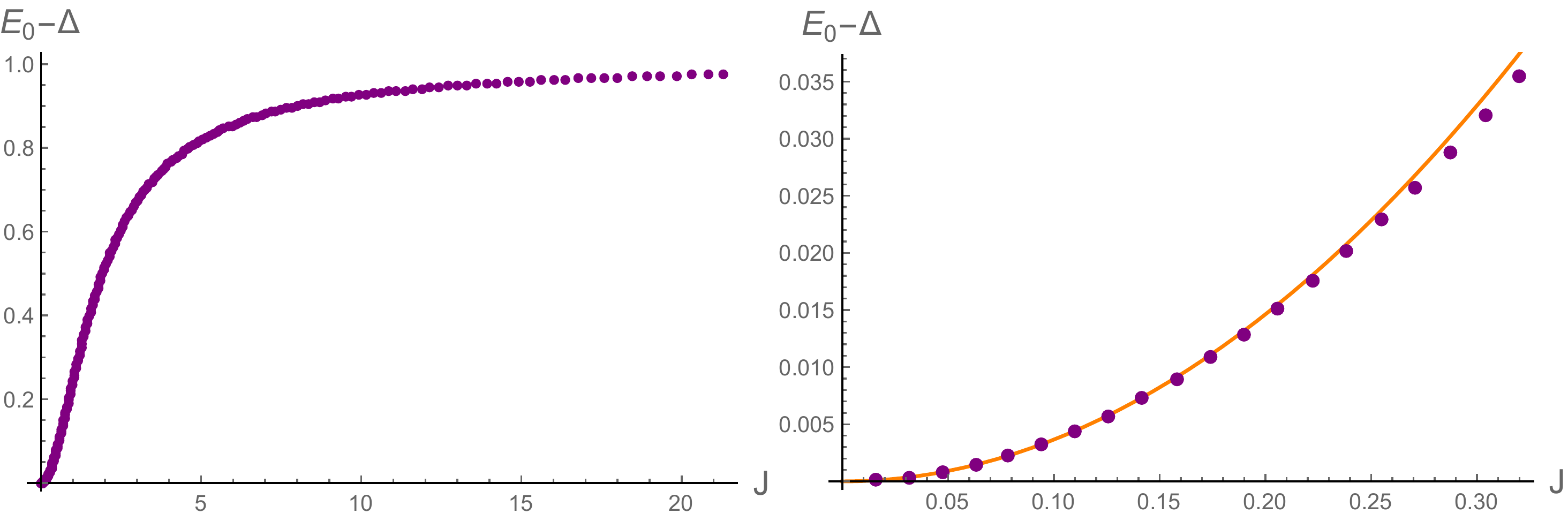}
\end{subfigure}
\newline
\end{figure}

 In practice we study the simplest setup to motivate our point. We consider a relevant scalar ($m^2=-2$) perturbation in global AdS4 background. The boundary of the spacetime is $S^2\times R$.  The scalar field has the near-boundary asymptotic expansion 
 \beq
 \phi = \frac{J(t)}{r} + \frac{\epsilon(t)}{r^2} + \ldots, \label{asymptotic_exp_scalar}
 \eeq   
 We turn on a time independent $J$ (and also $\epsilon$ in Section \ref{sec:alternative}). In the Section \ref{sec:model} we construct the $T=0$ solution , which has been christened AdSJ\footnote{We intentionally avoid any use of the word "soliton" to avoid confusion with other bosonic holographic configurations\cite{Gentle_2012} and AdS-solitons\cite{PhysRevD.59.026005}. }. We prove a version of C-theorem and discuss the consequences in the Section \ref{sec:C}. In the Section \ref{sec:HP}, we study the BH solution in AdSJ geometry and the Hawking-Page transition numerically. We also study a systematic perturbative computation of $T_{HP}$, in $J$ and match the perturbative result with our exact numerics. This has been done in the Section \ref{sec:perturb}. We find broad similarity with Claims \ref{item:main}. In Appendix \ref{app:planar-limit}, we discuss the approach to planar limit. 
       
 Scalar condensates, i.e. BH solutions with scalar hair have been discussed in works related to Holographic superconductors \cite{Hartnoll:2008kx} as well as for global AdS \cite{Basu:2010uz,Dias_2012}. A scalar condensation induced by another scalar has been discussed in \cite{Basu:2019pxw}. However, investigating any kind of  second order phase transition \cite{Hartnoll:2016apf} is not the focus of this work. The study of boson stars \cite{2003CQGra..20R.301S}, a bosonic lump in space time, is more than half a century old and was pioneered by Wheeler and Kaup \cite{PhysRev.97.511,1968PhRv..172.1331K}. Boson star like solutions are also constructed in AdS \cite{Astefanesei:2003qy}. They are prevented from collapse by gravitation and/or by a boundary chemical potential corresponding to a global charge. Our solution is different and could be thought of as turning on a relevant operator in the boundary. As it is a relevant deformation, in some sense our AdSJ solution has a infinite negative mass.

\section{Model \& equations of motion and AdSJ solution}\label{sec:model}
 
 We consider four-dimensional Einstein gravity with a negative cosmological constant $\Lambda$ and a massive scalar field $\phi$. The action is 
 \beq
 S = S_{EH} + S_{\phi} = \frac{1}{16 \pi G} \int d^{d+1} x \sqrt{-g} (R - 2 \Lambda) - \int d^{d+1} x \sqrt{-g} \left[\frac{1}{2}(\partial \phi)^2 + V(\phi) \right], \label{action}
 \eeq
 where the scalar field potential is given by $V(\phi) = \frac{1}{2} m^2 \phi^2$, with $m^2 = -2$. Unless stated otherwise in a explicit manner, we are working in units such that $4 \pi G = 1$ and $\Lambda = - 3/L^2$, $L=1$. We are considering spherically symmetric solutions. Our metric ansatz is 
 \beq
 ds^2 = - F e^{-2 \delta} dt^2 + F^{-1}dr^2 + r^2 d\Omega_{d-1}^2, \label{ds2}
 \eeq
 where $\delta$, $F$ and the scalar field $\phi$ are functions of $t$ and $r$. The boundary of the AdS$^{d+1}$ lies at $r=\infty$ and the center is at $r=0$. We will confine ourselves to $d=3$. For our numerics we often use a co-ordinate redifination $\tan(x)=r$ (see Appendix \ref{app:tanx}), which maps $r$ to a compact domain $x\in [0,\pi/2]$.
 
  A useful field redefinition is given by
 \beqa
 \Phi(t,r) = \phi'(t,r) \ , \qquad \Pi(t,r) = \frac{e^{\delta(t,r)}}{F(t,r)} \dot \phi(t,r)\, .
 \label{PhiPi}
 \eeqa
 Upon this redefinition, the scalar field equations of motion can be cast in the form
 \beqa
 \dot\Phi &=& \left(F e^{-\delta}  \Pi \right)'  \ , \\
 \dot\Pi &=&   \displaystyle \frac{1}{r^2}\left(r^2 F e^{-\delta} \Phi\right)' + 2 e^{-\delta} \phi \, . 
 \eeqa
 From the Einstein equations we obtain
 \beqa
 F' &=&\displaystyle \frac{-F+3 r^2+1}{r}-r F \left(\Pi^2+\Phi^2\right)+2 r \phi^2,\\
 \delta' &=&-\displaystyle   r \left( \Phi^2 + \Pi^2 \right) \, .  
 \eeqa
 For the static solutions relevant for the computation of the phase diagram, the equations reduce to 
 \beqa\label{eq:eqns}
 &&\delta '(r)+r \phi '(r)^2 = 0, \nonumber  \\
 &&F'(r)-F(r) \left(-r \phi '(r)^2-\frac{1}{r}\right)-2 r \phi (r)^2-3 r-\frac{1}{r}=0, \nonumber\\
 &&\phi ''(r)-\frac{\left(-F(r)-2 r^2 \phi (r)^2-3 r^2-1\right) \phi '(r)-2 r \phi (r)}{r F(r)}= 0. 
 \eeqa
 The scalar field has the near-boundary asymptotic expansion 
 \beq
 \phi = \frac{J(t)}{r} + \frac{\epsilon(t)}{r^2} + \ldots, \label{asymptotic_scalar}
 \eeq
 where $r$ is a radial Schwarzschild coordinate. 
 In the standard quantization of the scalar field, $J$ corresponds to the source of the scalar operator $\mathcal O$ of conformal dimension $\Delta = 2$ dual to $\phi$, while $\epsilon$ determines its vacuum expectation value; for the alternate quantization, the roles played by $J$ and $\epsilon$ get reversed.

The metric behaves as 
\beqa
&&F(r) =r^2+\left(J^2+1\right)-\frac{m_0}{r}+\frac{2 \epsilon ^2+2 J^4+J^2}{r^2}+ O\left(r^{-3}\right), \label{asymptotic_exp_f}\\
&&\delta(r) = \frac{J^2}{2 r^2}+O\left(r^{-3}\right), \label{asymptotic_exp_d}
\eeqa
where we have chosen the gauge $\delta(\infty) = 0$, which makes the coordinate $t$ to agree with the proper time of a static boundary observer.  In our case, we are interested in determining how the introduction of a finite, time-independent and homogeneous source modifies the phase structure of black holes in asymptotic global AdS.

\subsection{AdSJ solution}
At the probe limit (small $J$), where the scalar amplitude is small and is gravitationally decoupled, we have (with $x=\tan^{-1}(r)$, see Appendix \ref{app:tanx}),

\begin{align}
\phi(x)=x \cot(x).
\end{align}

\begin{figure}[h!]
	\begin{center}
		\begin{subfigure}{5cm}
		\includegraphics[width=6cm]{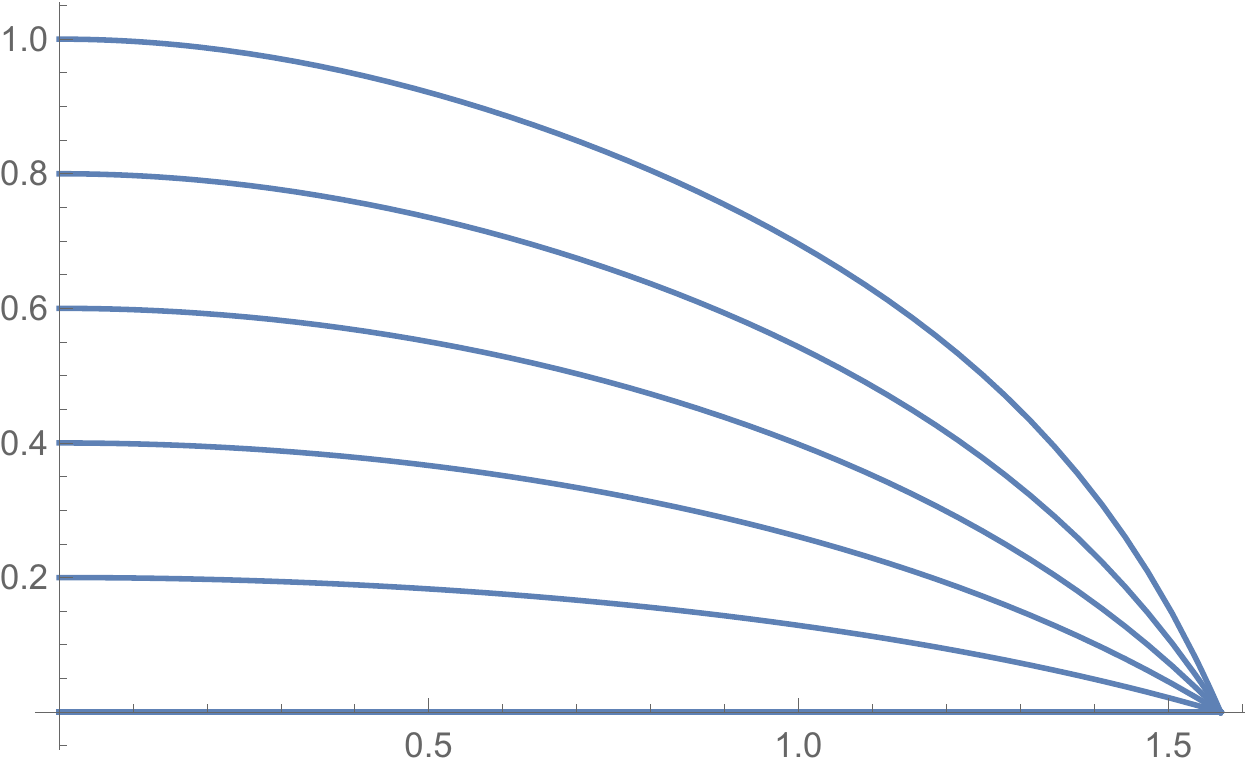}
		\includegraphics[width=6cm]{figs/Rs.pdf}
	\caption{Upper: Scalar field profile with $x$. Lower: $R$ vs $x$ plots.  $J=0.3, 0.7, 1.1, 1.7, 2.5$, increasing away from the $x$-axis. }	
	\end{subfigure}
	\hspace*{2cm}
	    \begin{subfigure}{5cm}
		\includegraphics[width=6cm]{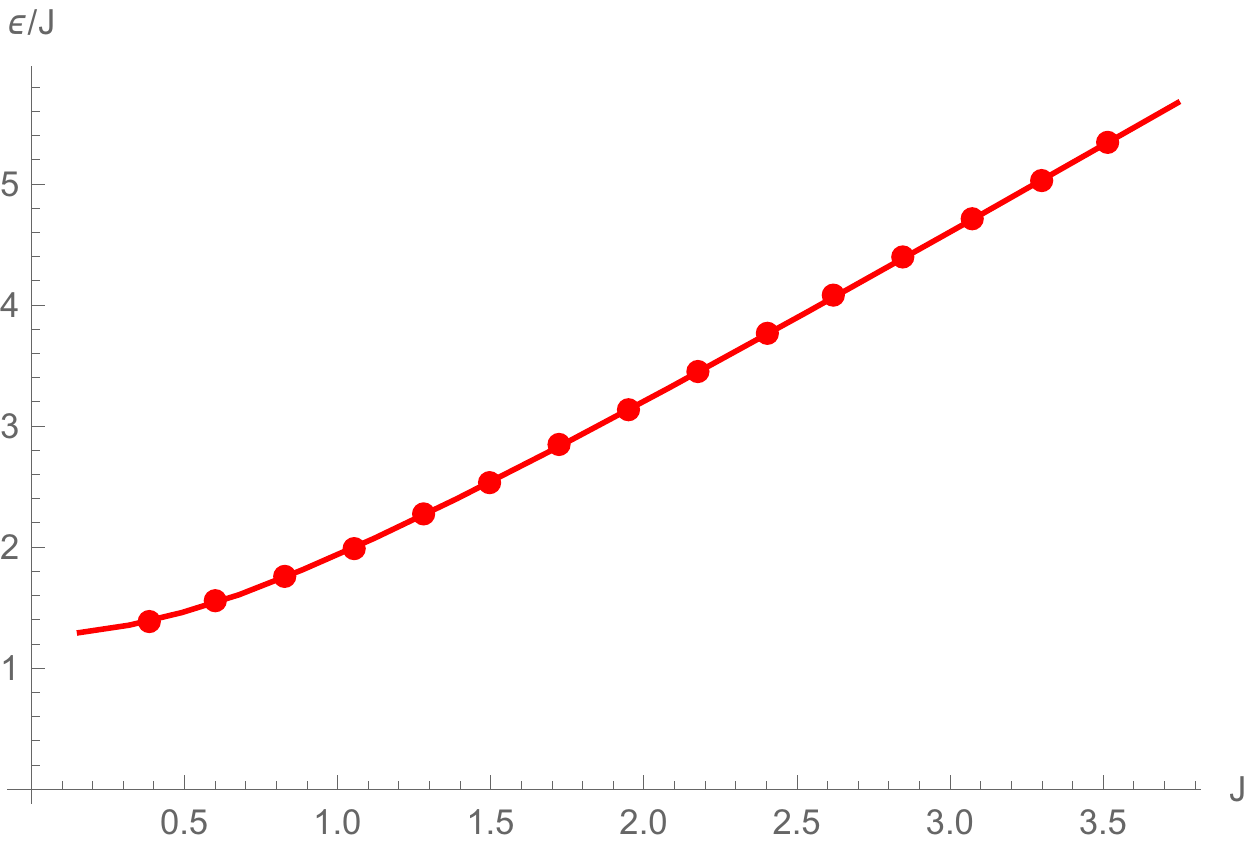}
		\hspace*{-0.57cm}
		\includegraphics[width=6cm]{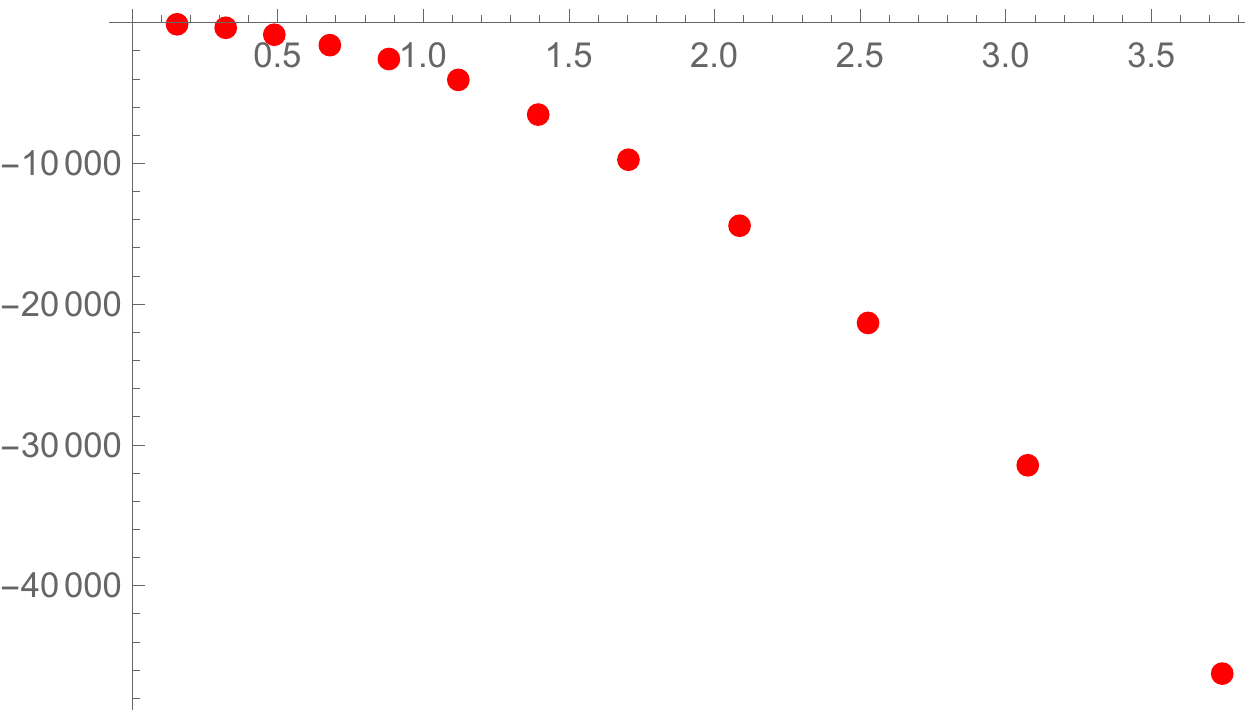}
		\caption{Upper: $\epsilon/J$ vs $J$. Lower: Volume of the space vs $J$.}
		\end{subfigure}
	\end{center}
\caption{AdSJ solution and its properties.}\label{fig:AdSJ}
\end{figure}

That gives $\epsilon/J=2/\pi$.  The fully non-linear solutions for higher values of $J$ has been shown in Figure \ref{fig:AdSJ}. We notice that the with increasing $J$ the scalar curvature ($R$) gets  more negative and its amplitude increases inside the bulk.

 \subsection{C-theorems on a compact space}\label{sec:C}

The holographic c-theorem(or it's higher dimensional counterpart a-theorem) naturally appears in the context of a boundary deformation. In field theoretic context c-theorem implies that certain quantity (central charge) of a CFT increases monotonically under RG flow, whereas in holography it implies that there exists a monotonically varying function in the bulk, from the UV region to the IR.  Holographic c-theorems are often discussed when the boundary is flat. However in our case the  the boundary is $S^3 \times S^1$, where the spatial part $S^3$ is a compact manifold with a finite volume. Compact manifold provides a natural IR cutoff to the RG flow, and RG flow may not continue to IR fixed point. We  argue instead that c-theorem like arguments may be used to prove that a deformation in the boundary reduces the bulk volume. As we soon discuss, this has an interesting physical and geometrical consequence. 

The proof of this theorem relies on the Null Energy condition, and uses the Einstein equations. Following in the footsteps of earlier calculations, we can quickly show that such a function exists in our case. \\
Consider the difference between $G_t^t$ and $G_r^r$, the Einstein tensors for the metric, which for this ansatz \ref{ds2}, equals to $\frac{2F(r)\delta^{\prime}(r)}{r}$. Using Einstein equations, this can be equated to
\begin{equation}
\frac{2F(r)\delta^{\prime}(r)}{r}=G_t^t-G_r^r=8\pi G(T_t^t -T_r^r) 
\end{equation}
By the null energy condition, the right hand side is always non-positive. Hence, we can conclude that 
\begin{equation*}
\frac{2F(r)\delta^{\prime}(r)}{r}<0
\end{equation*}
Since the emblackening factor $F(r)$ is always positive in the region we're interested in(outside the horizon for a Black-Hole and everywhere in the AdSJ case), it shows that 
\begin{equation}\label{eq:delta}
\delta^{\prime}(r)<0
\end{equation}
We have now shown that a function exists whose radial derivative is always negative in the bulk.

This is, however, is not the only conclusion. If we look at the scalar field equation,
\begin{align}\label{eq:phi} 
\partial_r(\sqrt{g} g^{rr}\partial_r \phi)=\sqrt{g} V'(\phi)
\end{align}
the above equation leads to an attractor like mechanism \cite{Goldstein_2005}. For our case, the attractor mechanism says that $\phi^2$ is an monotonically decreasing function of the radial co-ordinate, i.e. 
\begin{align}\label{eq:monphi}
\phi \phi'<0
\end{align}. 
This monotonicity of $\phi^2$ follows from the fact that the scalar field $\phi$ and its double derivative with the radial coordinate (or any other monotonic reparametrization of $r$) has the same sign. Unlike attractor mechanism in flat space, $\phi$ reaches a finite value at the center of the space which implies $\phi'(0)=0$ from smoothness.  

	Let's look at the emblackening function $F(r)$. From \eqref{eq:eqns}, we get,
\begin{align}\label{eq:black1}
F'(r)-F(r) \left(\delta'-\frac{1}{r}\right)-2 r \phi (r)^2-3 r-\frac{1}{r}=0 \\
\Rightarrow F(r)=\frac{1}{r}e^{\delta(r)}\int_{0}^{r} e^{-\delta(r)} (2 r^2 \phi^2+3 r^2+1) dr
\end{align}
The above expression is a quantification of the intuitive claim that turning on $J$ augments the cosmological constant and dumps negative energy in the bulk. With a rearrangement of  \eqref{eq:eqns} we also have an alternative expression for the emblackening function:
\begin{align}\label{eq:black2}
F(r)&=1+r^2+\frac{1}{r  }\int_{0}^{r} -e^{\delta}\partial_r(r^2 F e^{-\delta} \phi \phi ') d r \\
&=1+r^2-r F  \phi \phi '+\frac{1}{r  }\int_{0}^{r} \delta'r^2 F \phi \phi ' d r \\
\end{align}
From \eqref{eq:black1} and \eqref{eq:black2} we may prove the following inequality,
\begin{align}\label{eq:inequalityF}
1+r^2<F(r)<1+r^2+\frac{1}{r}\int_0^r 2 r^2 \phi^2 dr
\end{align}
giving us an idea how much $J$ affects the emblackening factor. In proving the above inequality we have used \eqref{eq:monphi} and the monotonicity condition  \eqref{eq:monphi}.

\subsection{Volume of the space and EE}
	This helps us in making another important conclusion. The volume element, $e^{-\delta(r)}r^2\sin{\theta}$, from which we can write down the total (regularized and compared to global AdS4) volume of the spacetime to be
\begin{equation*}
Vol=\int  (e^{-\delta(r)}-1)r^2\sin{\theta}dt dr d\Omega 
\end{equation*}

	This volume decreases with the introduction of the scalar field as $e^{-\delta(r)}<1$. 

	Another quantity of interest is the EE. We would concentrate on calculating the Entanglement Entropy of a circular area in the boundary. Boundary is an $S^2 \times R$ here. If we take an area ($A$) enclosed by a curve $\cal C$, then according to Ryu-Takyanagi \cite{Ryu_2006} EE is the area of a  maximal $d-2(=2)$ co-surface in AdSJ which ends on $\cal C$. For an area in boundary enclosed by a  great circle ($A_{max}$),  we have, 
\begin{align}
EE_{A_{max}}&= 2 \pi \int_0^{r_o} \frac{r}{\sqrt(F(r))}  \,dr \\
\end{align} 
$F(r)=1+r^2$ for AdS4 and we have,
\begin{align}
EE^{AdS4}_{A_{max}}& \approx  2 \pi (r-1)
\end{align} 
The divergent first term is the "area" contribution and the second term is ${\cal O}(1)$ negative contribution \cite{BAKAS2015440}. For AdSJ we have,
\begin{align}
EE_{A_{max}}&= 2 \pi \int_0^{r_o} \frac{r}{\sqrt(F(r))}  \,dr = 2 \pi (r-\delta_{EE}(J))\\
\end{align} 
From \eqref{eq:inequalityF} we get,
\begin{align}
EE^{AdSJ}_{A_{max}} &<EE^{AdS4}_{A_{max}} \\
\Rightarrow \delta_{EE}(J)&>1\\.
\end{align} 
The ${\cal O}(1)$ contribution also seems to be increasing with $J$ from our numerics (see ). 
\begin{figure}[H]\label{fig:EE}
	\begin{center}
		\includegraphics[scale=0.5]{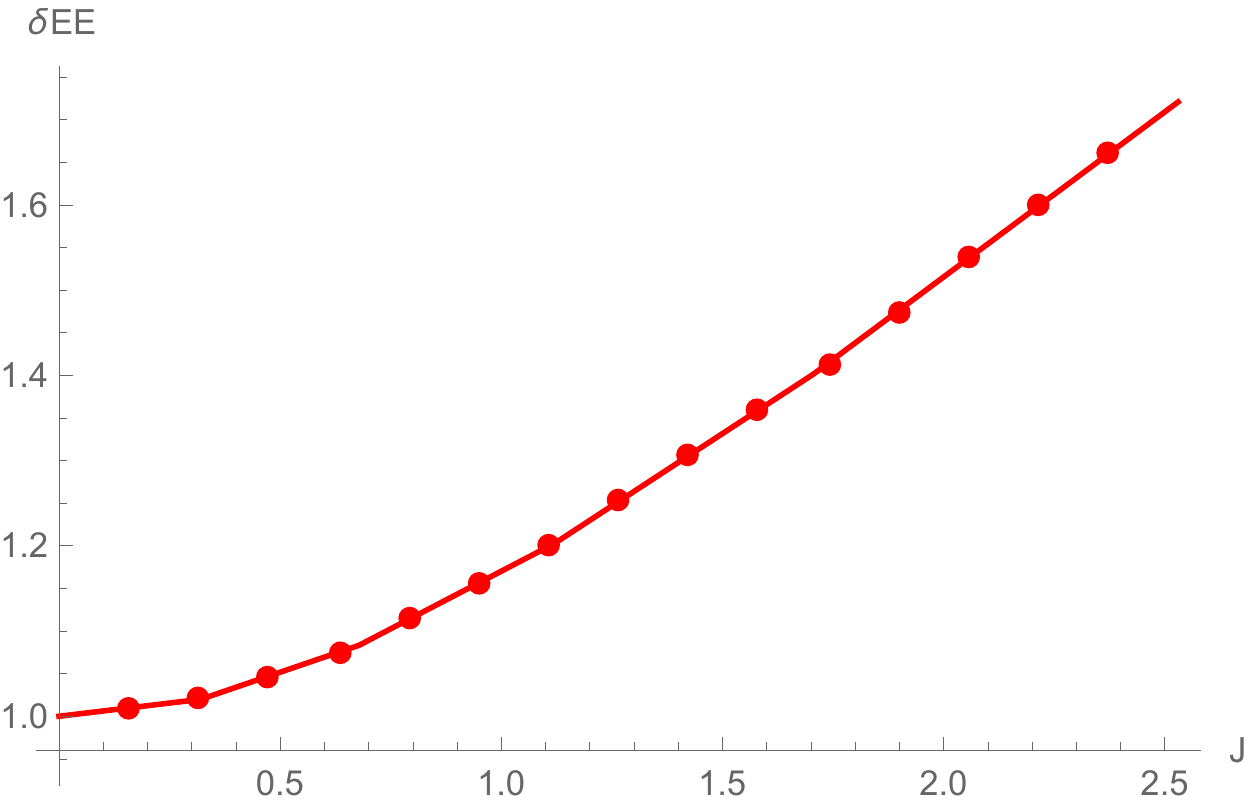}
	\end{center}
	\caption{A plot of $\delta_{EE}$ vs $J$.}
\end{figure} 
The extremal surface for EE calculation will venture more into the bulk as the size of the boundary region increases. Now because of the inequality \eqref{eq:inequalityF}, and the fact $F$ differs more from the $AdS4$ value of $f(r)=1+r^2$ as we go inside the bulk, we expect $\delta_{EE}$ to increase in magnitude with the size of the boundary region (see Claim \ref{item:main}). 
\subsection{Normal modes of the AdSJ solution}

The purpose of this section is to discuss the computation of normal modes of the AdSJ solution. We decompose the scalar field and metric as (using co-ordintes defined in Appendix \ref{app:tanx}),
\beqa
&&\phi(t,x) = \phi_s(x) + \gamma \cos {\omega t}\, \phi_1(x), \\
&&f(t,x) = f_s(x) + \gamma \cos {\omega t}\, f_1(x), \\
&&\delta(t,x) = \delta_s(x) + \gamma \cos {\omega t}\, \delta_1(x). 
\eeqa
where the subscript s refers to the background AdSJ solution, and $\omega$ is the frequency of the normal mode. By expanding the equations of motion to first order in the mode amplitude $\gamma$, we get the equations of motion for these fluctuations.  The reason for choosing this particular ansatz is that the overall time dependence in these equations factorizes. 

\begin{figure}[h!]
	\begin{center}
		\includegraphics[width=16cm]{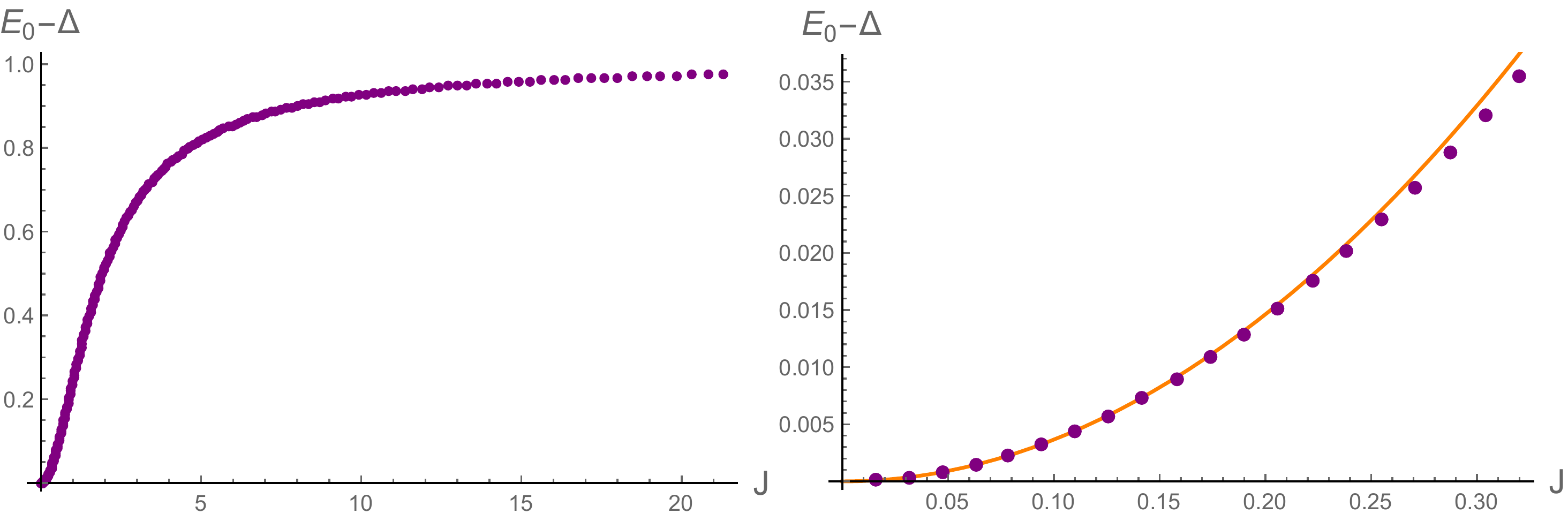}
		\caption{\small Left: Difference between the lowest eigenfrequency of the AdSJ, $E_0$, and its value on the AdS$_4$ vacuum, $\omega_0 = \Delta = 2$, as a function of the source $J$ supporting the AdSJ background. Right: as in the left plot, but for smaller $J$. The solid orange line is the leading order dependence of $E_0$ on $J$ predicted by perturbation theory.}
		\label{fig:eigenfrequency}
	\end{center}
\end{figure}

It is important to realize that the momentum constraint equation becomes, at first order, an algebraic relation that is solved by 
\beq
f_1(x) = - 2 \cos x \sin x \, \phi_s'(x) \phi(x). 
\eeq
In the same way, from the equation of motion for $\delta$ we get that 
\beq
\delta_1'(x) = -2 \cos x \sin x \, \phi_s'(x) \phi_1'(x). 
\eeq
Therefore, the only independent dof of the problem is $\phi_1(x)$. Its equation of motion follows from the scalar field's, once we take into account the relations between $f_1$, $\delta_1$ and $\phi_1$. The final result is a complicated but a linearized eigenvalue equation for $\phi_1(x)$,
\begin{align}
	&\phi_1''(x) + F_1(x) \phi_1'(x) + (\omega^2F_3(x)+ F_2(x)) \phi_1(x) = 0.\label{eq_normal_mode} \\
	\nonumber & F_1(x)=\frac{-\cot x + 2 \csc x \sec x + \tan x + \cos 2x \csc x \sec x f_s(x) + 2 \tan x \, \phi_s(x)^2}{f_s(x)} \\
	\nonumber & F_2(x)= \frac{  2 \left(\sec^2 x + 4 \tan x\, \phi_s(x) \phi_s'(x) - (\cos 2 x - 2) \phi_s'(x)^2 + 2 \sin^2 x \, \phi_s(x)^2 \phi_s'(x)^2\right)}{f_s(x)} \\
	\nonumber &F_3(x)=\frac{e^{2 \delta_s(x) }}{f_s(x)^2}
\end{align}
The value of frequency $\omega$  is determined by imposing the following boundary conditions: 
\begin{itemize}
	\item Smoothness at the origin, i.e., $\phi_1'(0) =0$. 
	\item Normalizability. The source $J$ that supports the AdSJ is kept invariant by the perturbation, i.e., $\phi_1'(\pi/2) = 0$. 
\end{itemize}
\noindent The numerical values of the lowest eigenfrequency are plotted in Fig.\ref{fig:eigenfrequency}a. We observe that at large $J$, the eigenfrequency saturates to a constant value. On the other hand, at small $J$ it reduces to its value on the AdS$_4$ background, plus a positive quadratic correction. 

The value of this quadratic correction can be computed perturbatively, by expanding $\phi_1(x) = \phi_{1,0}(x) + J^2 \phi_{1,2}(x)$, $\omega = 2 + \omega_2 J^2$ and introducing these expressions with \eqref{sol_expansion_1}-\eqref{sol_expansion_3} and \eqref{sol_solution_1}-\eqref{sol_solution_3} into \eqref{eq_normal_mode}, expanding them to order $J^2$. First, we find that $\phi_{1,0}$ is given by the lowest normal mode of the scalar field  on the AdS$_4$ bakground,  
\beq
\phi_{1,0}(x) = \cos^2 x, 
\eeq
as expected. 
The crucial point is that, at the next order, the requirement that $\phi_{1,2}$ is simultaneously normalizable and regular can only be met if the second order correction to the frequency, $\omega_2$, takes the  value 
\beq
\omega_2 = 3 - \frac{26}{\pi^2} = 0.365649 > 0. 
\eeq
In Fig.\ref{fig:eigenfrequency}b, we plot the numerical results for the lowest eigenfrequency (purple dots) vs this perturbative prediction (solid orange). It is clear that at small $J$ both computations agree very well. 

It is natural to interpret the value of $E_0$ as the mass gap of the dual field theory, in the sense that $E_0$ sets the rest energy of the lighest scalar excitation. Heuristically, we might assume that, since the typical size of this excitation should scale as $1/E_0$, the deconfinement temperature must also be proportional to $E_0$, as we would need a medium of thermal wavelength $\epsilon \sim 1/E_0$ to resolve the size of the scalar glueball and melt it. If this expectation is correct, the fact that $E_0$ tends to a constant in the large $J$ limit leads to the prediction that the Hawking-Page temperature should also saturate in this limit.\footnote{We have already established that there is no Hawking-Page transition in the planar limit. This implies that the phase transition temperature scales as $T_{HP} \sim J^\alpha$, with $\alpha < 1$. The reasoning of this paragraph is equivalent to the statement that $\alpha = 0$.}$^,$

\subsubsection{Massless scalar field modes on the AdSJ background}

The purpose of this section is to compute the normal modes of a probe massless scalar field $\psi$ on the AdSJ background. The underlying idea is interpreting these normal modes as a proxy for the mesonic excitations in the system. Note that $\psi$ is different from the scalar field $\phi$ that sources the  geometry. Given that $\psi$ is a probe field, its equation of motion on the AdSJ background is given by 
\beq
\partial_a \left(\sqrt{-g_s}g_s^{ab} \partial_b \psi \right) = 0,
\eeq
where $g_s$ is the AdSJ metric. If we assume that $\psi$ is spherically symmetric and decompose  $\psi(t,x) = \hat\psi(x) e^{-i \omega t}$, we get  (using the co-ordintes defined in Appendix \ref{app:tanx})
\beq
\hat\psi''(x) + \left(2 \csc x \sec x + \frac{f_s'(x)}{f_s(x)} - \delta_s'(x) \right) \hat \psi'(x) + \frac{e^{2 \delta_s(x)}\omega^2}{f_s(x)^2} \hat \psi(x) = 0, \label{psi_eq}
\eeq
where the subscript $s$ refers to quantities evaluated on the AdSJ background. Equation \eqref{psi_eq} is to be solved demanding that $\hat \psi(x)$ is regular at $x = 0$, $\lim_{x \to 0} x \hat \psi(x) = 0$, and normalizable, $\hat \psi(x = \pi/2) = 0$. Provided that these boundary conditions are imposed, $\omega$ corresponds to a normal mode frequency.

\begin{figure}[h!]
	\begin{center}
		\includegraphics[width=11cm]{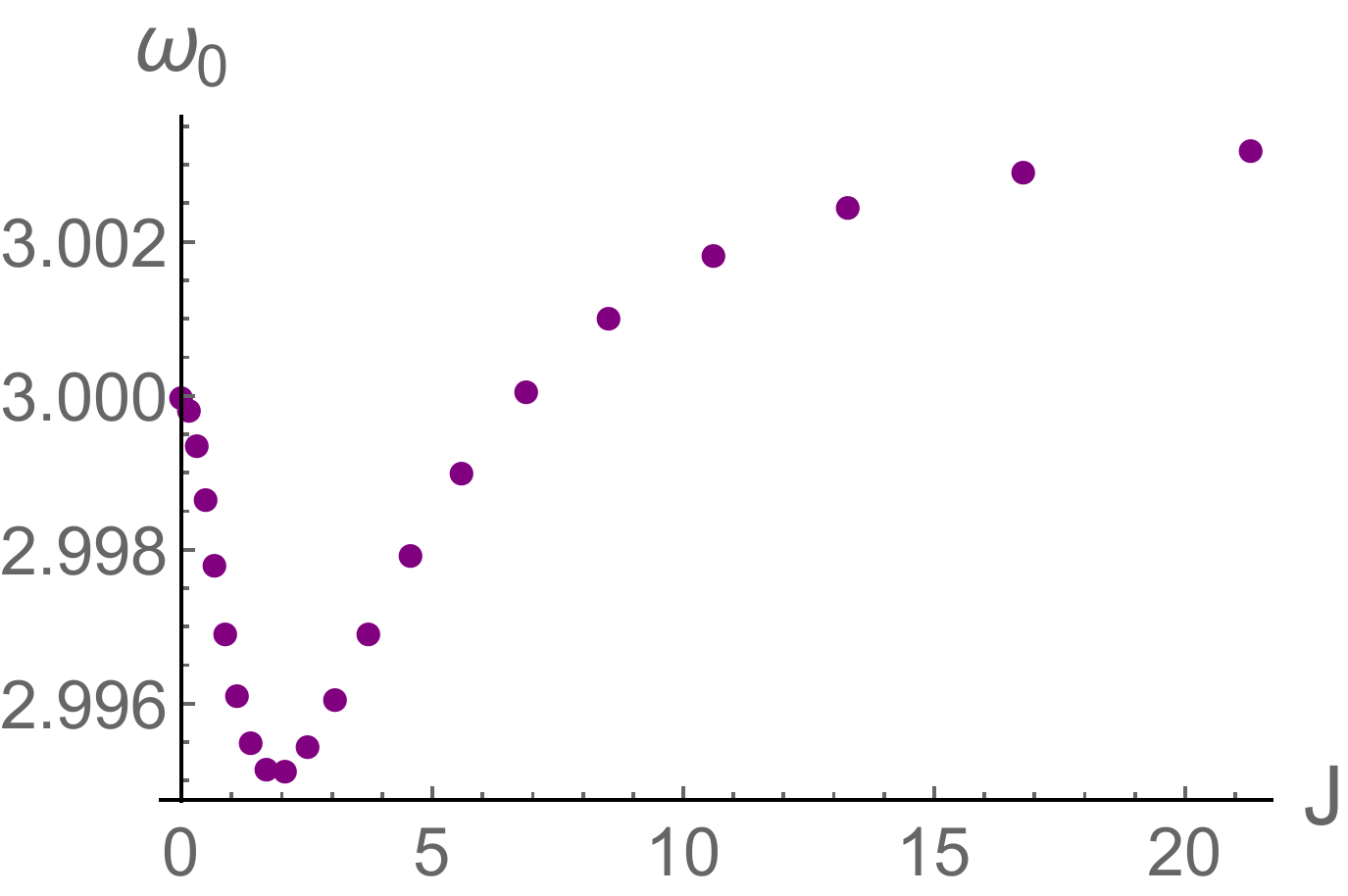}
		\caption{\small Fundamental mode of a probe massless scalar field on the AdSJ background.}
		\label{fig:psi_sound}
	\end{center}
\end{figure}

We have solved \eqref{psi_eq} numerically, employing a pseudospectral algorithm. Since neither \eqref{psi_eq} nor the boundary conditions we are imposing distinguish between the standard and the alternative quantization for $\psi$, we only need to solve the problem once for each AdSJ solution. In Fig.\,\ref{fig:psi_sound} we plot the frequency of the lowest normal mode, $\omega_0$, against $J$ (the plot in terms of $\epsilon$ is qualitatively similar). It is manifest that the behavior of the lowest eigenfrequency is not correlated at all with the behavior of the Hawking-Page temperature, neither in the standard nor in the alternative quantization.

\section{Deconfinement and BHs at finite J}\label{sec:HP}

\begin{figure}[h!]
	\begin{center}
		\includegraphics[width=12cm]{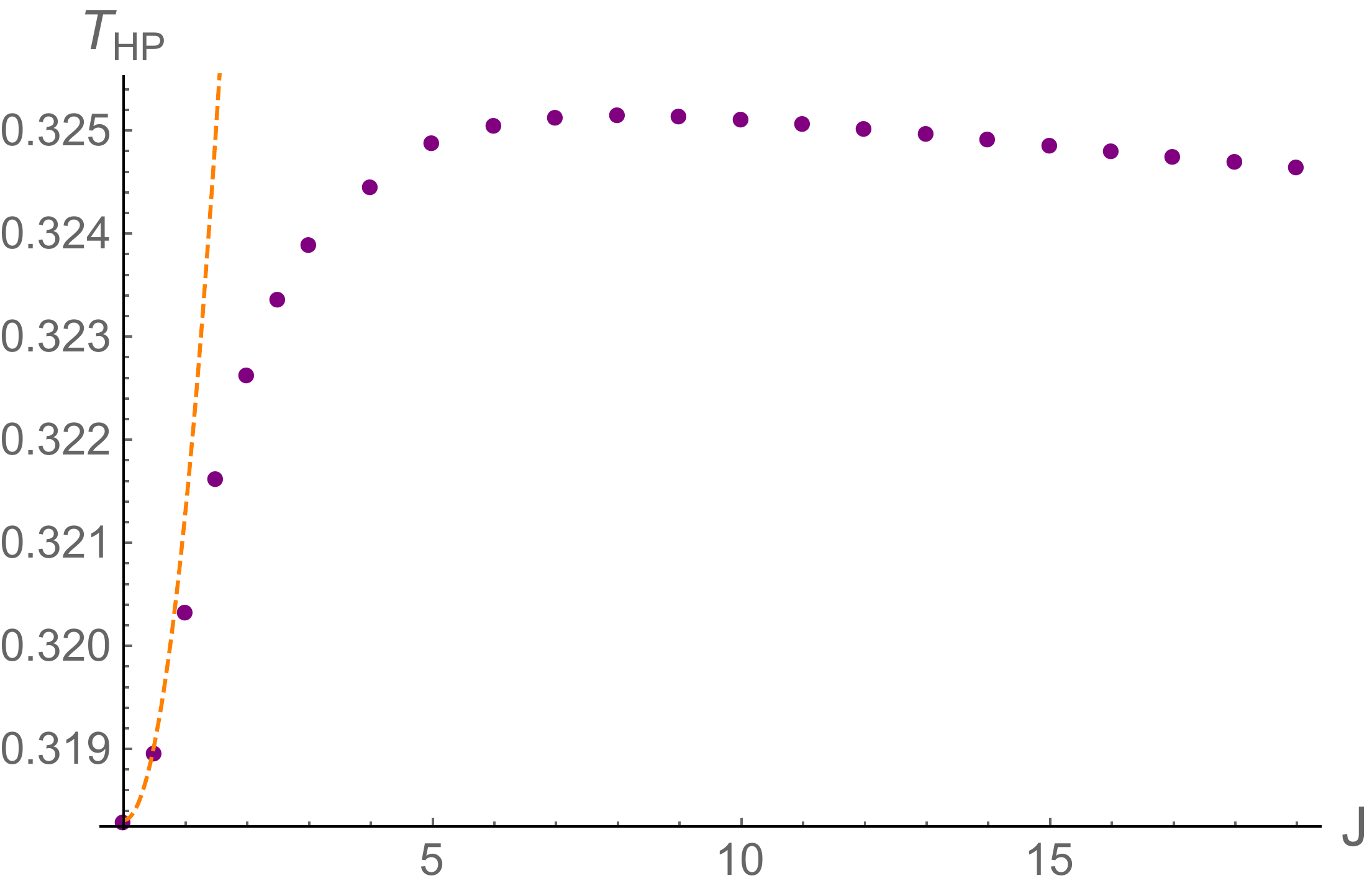}
		\caption{\small Phase transition temperature $T_{HP}$ vs relevant coupling $J$ (purple dots). The leading order perturbative prediction for $T_{HP}$ (eqn. \eqref{THP_corr_1}) corresponds to the dashed orange line.}
		\label{fig:T_fullynonlinear}
	\end{center}
\end{figure}
In this section we undertake the task of determining the grand-canonical phase diagram of our theory at finite $J$ and $T$.  At $J=0$, the well known story of Hawking-Page transition \cite{hp,Witten:1998zw} is the following. At $T=0$ there exists only one semi-classical solution which is global AdS. As the temperature is increased, at the nucleation point $T=T_N$, a new solution of big and small black holes become possible classical saddles. At the nucleation point both big and small black holes have the same  mass and size. As temperature is increased further, the area(also the entropy) of the big black hole increases and the the opposite happens for small black holes. At $T=T_{HP}$ the free energy of the big BH becomes more than global AdS and the it becomes the dominating saddle. This is identified with deconfinement transition in the boundary theory \cite{Witten:1998zw}. Possibly, there exist a few other gravity solutions including BH solutions, and even at a higher temperature transitions like hegadorn transition are possible \cite{Aharony_2004}. We won't expound upon these directions in this work.

 \subsection{The free energy computation at finite $J$}

In order to find out which solution dominates the grandcanonical partition function at a given point in the $(J, T)$-plane, we need to calculate their Gibbs free energy. 

As usual, the free energy $F$ is given by the renormalized Euclidean on-shell action of the solution as $F = T S_{E}$. To compute this quantity, we have to add a series of countertems to the action \eqref{action} in such a way that, for our choice of boundary conditions, $S_E$ is stationary on the saddle points we are interested in. This is equivalent to the statement that the variational problem is well-defined; as a byproduct, the resulting action comes out finite. 

The full computation of $S_E$ can be found in appendix \ref{app:global}. We quote below the final result for the free energy density
\beq
f \equiv \frac{F}{\textrm{Vol}(S^2)} = -\frac{m_0}{16 \pi G} - J \epsilon  + \frac{r_{IR}}{8 \pi G} + \int_{r_{IR}}^\infty dr \frac{1-e^{-\delta(r)}}{8 \pi G}, \label{f_x}
\eeq 
where $r_{IR} = 0, r_h$ for AdSJ and black holes respectively. 

\subsection{Results}

\begin{figure}[t!]
	\begin{center}
		\includegraphics[width=11cm]{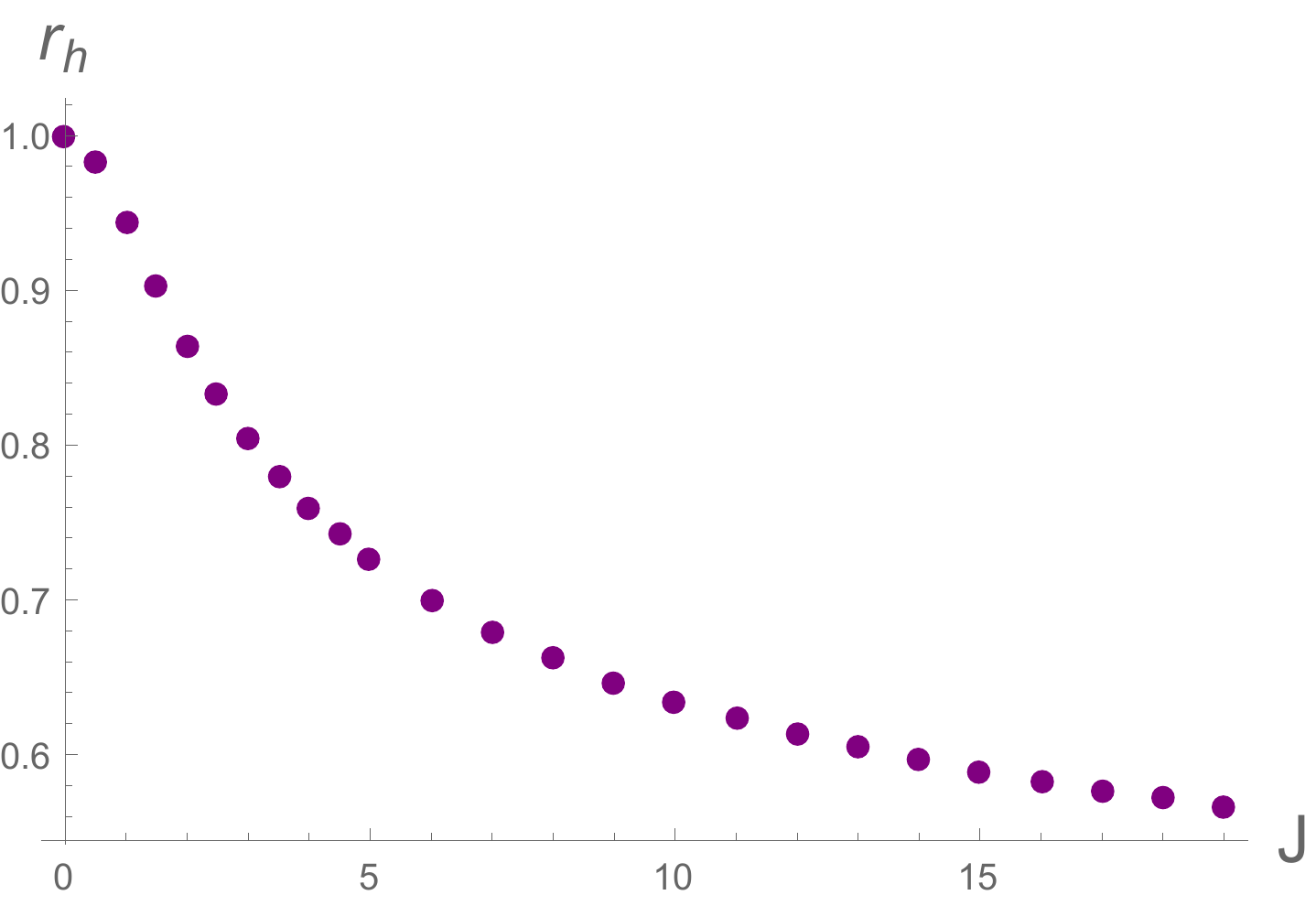}
		\caption{\small Radius of the first thermodynamically dominant black hole in terms of the coupling.}
		\label{fig:radius}
	\end{center}
\end{figure}

The fundamental qualitative story of the HP transition remains almost the same at finite $J$.  At $T=0$ there exists only one classical solution which is global AdSJ with no horizon but a nontrivial scalar field profile. At higher temperatures, there is another competing semiclassical saddle: the big hairy BH, which becomes dominant at a coupling dependent transition temperature $T_{HP}(J)$. However, the coupling $J$ has a significant influence on the entropy difference at the phase transition: as $J$ increases, this entropy difference goes down. In order to illustrate this, in Fig.\,\ref{fig:radius} we plot how the radius of the first thermodynamically dominant black hole changes with the coupling. Its decrease is manifest. This monotonic decrease is enough to overcome the small variations of $T_{HP}$. In Fig.\,\ref{fig:T_fullynonlinear}, we plot the Hawking-Page temperature as a function of the relevant coupling $J$. We note that $T_{HP}$ is monotonic for small enough $J$ and we have demonstrated that with a perturbative analysis. We also see that $T_{HP}(J)>T_{HP}(0)$ for all $J$. As discussed in Appendix \ref{app:planar-limit}, this last behavior is compatible with the absence of a planar limit for the Hawking-Page transition. Unfortunately, we have been unable to determine beyond any trace of doubt whether $T_{HP}$ approaches zero as $J \to \infty$ or saturates to a finite value. 

 It is manifest that $T_{HP}$ is not globally monotonic:  there is a maximum temperature $T_{HP}^{*} = 0.3252$, which is attained at $J^* = 8.124$. Past this point, $T_{HP}$ decreases. It must be noted that, overall, the relevant coupling has a small influence on the value of $T_{HP}$. Our numerical results strongly suggest  that $T_{HP}$ never exceeds $T_{HP}^*$ and as we discussed $T_{HP}$ never goes below it's zero temperature value $1/\pi$ for large $J$. If this is the case, the maximum relative variation of $T_{HP}$ with the coupling would be  given by $(T_{HP}^* - 1/\pi)/(1/\pi) = 2.155\,\%$.

\subsection*{Solution at large J}

This subsection sums up the numerical results we have found for $T_{HP}(J)$ at arbitrary coupling. To obtain reliable results, we are forced to be extremely precise in determining the free energy densities of the AdSJ and black hole solutions. To this end, we have solved equations \eqref{eqns} employing a pseudospectral method, checking our results by monitoring the agreement between expressions \eqref{f_x} and \eqref{smarr}.

\begin{figure}[h!]
	\begin{center}
		\includegraphics[width=16cm]{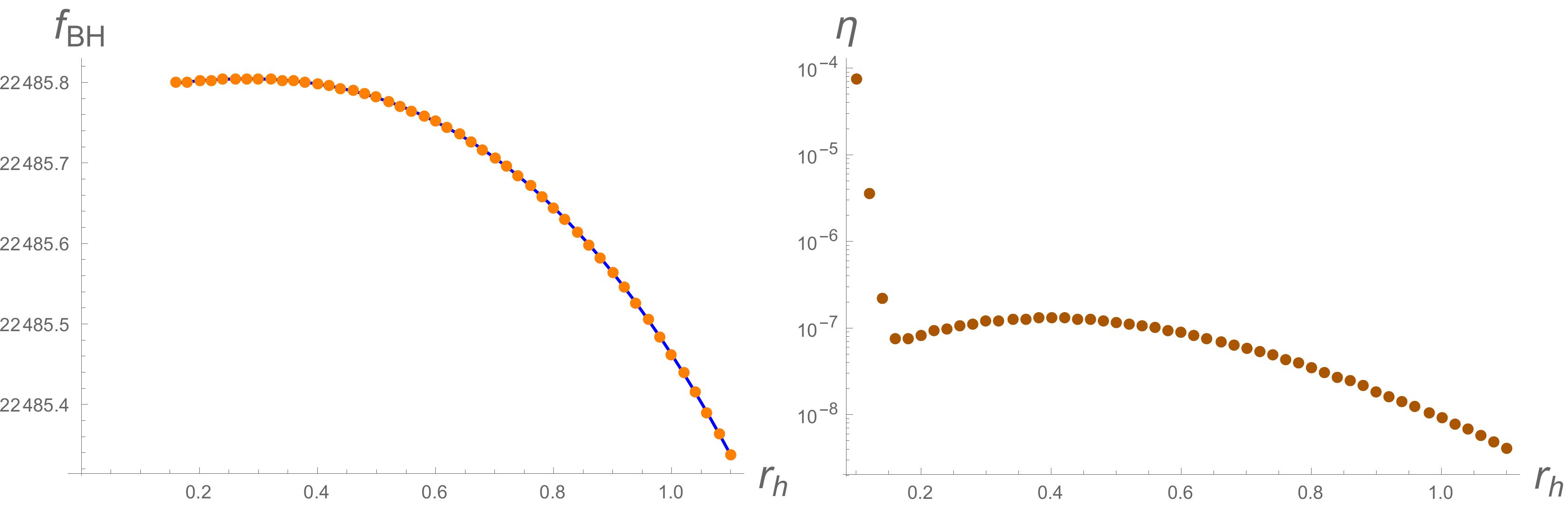}
		\caption{\small Left: At $J = 45$, free energy density of the black hole family as computed by eqn. \eqref{f_x} (orange dots) and by the Smarr relation \eqref{smarr} (solid blue). Right: The agreement of the results of the left plot, as measured by the relative error parameter $\eta$.}
		\label{fig:test}
	\end{center}
\end{figure} 

As an example of the power of the pseudospectral method we have employed, in Fig.\,\ref{fig:test} we consider the black hole family with $J = 45$ and compare expressions \eqref{f_x} and \eqref{smarr} for the free energy density. The grid we have employed for this computation consists of $121$ collocation points. We have introduced the quality parameter $\eta$, defined as
\beq
\eta = \frac{|f_1 - f_2|}{(f_1 + f_2)/2} 100,  
\eeq
where sub-indices $1$,$2$ refer to the free energy densities computed with each method. As we can see, they are in excellent agreement: at this coupling, the phase transition takes place at $r_h = 0.49682$; according to Fig.\,\ref{fig:test}, the relative error in the black hole free energy density at that radius is $O(10^{-7}\,\%)$.

\section{The perturbative computation of $T_{HP}(J)$}\label{sec:perturb}

It is desirable to check our numerical results against a (semi)analytic computation. To this end, we are going to show how the lowest order correction to $T_{HP}$ can be determined with almost no effort. Here we focus on $J<<1$ limit. We are going to compute this correction in two alternative ways. The first one requires only the knowledge of  the expression for the free energy derived in the previous section. The second one relies essentially on the First Law (it is essentially the derivation of the Clausius-Clapeyron relation for our problem). 

\subsection{First method}
\label{THP_pert_1}

The argument we employ is the following. We know that, at $J = 0$, the radius of the first thermodynamically dominant black hole is $r_c^0 = 1$, where
\beq
f_{BH}^0(r_c^0) = 0. 
\eeq
At finite $J$, this Schwarzschild black hole is replaced by a hairy black hole with the same horizon radius, while global AdS$_4$ is replaced by a AdSJ background. Therefore, the difference in their free energies gets modified as 
\beq
\Delta f = f_{BH}(r_c^0) - f_s \neq 0, 
\eeq
and the phase transition does not happen at $r_c = r_c^0$ anymore. 
At leading order, 
\beqa
&&f_{BH}(r_h) = f_{BH}^0(r_h) + J^2 \delta f_{BH}(r_h), \\
&&f_s = J^2 \delta f_s,  
\eeqa
and so the free energy difference is 
\beq
\Delta f = f_ {BH}^0(r_h) + J^2 (\delta f_{BH}(r_h) - \delta f_s) + O(J^4)
\eeq
The crucial observation is that, also at $O(J^2)$, we expect that the horizon radius at which the phase transition takes place changes as 
\beq
r_c = r_c^0 + J^2 \delta r_c = 1 + J^2 \delta r_c. 
\eeq
Substituting this expression into the free energy difference, we see that, at leading order,  it is sufficient to evaluate the change in the black hole free energy at $r_c^0$, since corrections coming from the change in $r_c$ only change this term at order $J^4$. Therefore, at the phase transition point, we have that 
\beq
\begin{split}
	\Delta f =& f_{BH}^0 (r_c^0 + J^2 \delta r_c) + J^2 (\delta f_{BH}(r_c^0) - \delta f_s) = f_{BH}^0 (r_c^0) +J^2 \left(\frac{df_{BH}^0}{dr} (r_c^0)  \delta r_c + \delta f_{BH}(r_c^0) - \delta f_s \right) \\
	=& J^2 \left( \frac{df_{BH}^0}{dr} (r_c^0)  \delta r_c + \delta f_{BH}(r_c^0) - \delta f_s \right) = 0, \label{x_c_correction} 
\end{split}
\eeq
up to terms $O(J^4)$. To determine the radius of the first thermodynamically dominant black hole at finite but small coupling, we just need: 
\begin{itemize}
	\item $f_{BH}^0$ as a function of $r_h$. 
	\item $\delta f_{BH}(r_c^0)$, which involves solving for the scalar field and metric fluctuations  over the Schwarzschild black hole of horizon radius $r_c^0 = 1$. 
	\item $\delta f_s$. 
\end{itemize}
The first and third quantities of the list above can be determined analytically; the second requieres  a straightforward numerical computation, since the linearized Klein-Gordon equation for the scalar field in the Schwarzschild black hole background is not exactly solvable (in global coordinates). 

The first quantity is easily shown to be equal to 
\beq
\frac{df_{BH}^0}{dr} (r_c^0) = \frac{1-3 r^2}{16 \pi  G}\bigg\rvert_{r_h = r_c^0=1}= - \frac{1}{8 \pi G} = -\frac{1}{2}
\eeq
in our units. On the other hand, in order to determine the AdSJ free energy, we need to find the AdSJ solution at $O(J^2)$. This is achieved by solving the linearized Klein-Gordon and Einstein equations over global AdS$_4$. We seek a time-independent, spherically symmetric solution supported by a nontrivial scalar field profile. Expanding in $J$
\beqa
&&\phi(r) = J \phi_1(r), \label{sol_expansion_1}\\
&&f(r) = (1 + r^2) + J^2 f_2(r), \label{sol_expansion_2}\\
&&\delta(r) = J^2 \delta_2(r), \label{sol_expansion_3} 
\eeqa
we obtain the following equations of motion for the fluctuations 
\beqa
&&\left(4 r^2+2\right) \phi _1'(r)+r \left(r^2+1\right) \phi _1''(r)+2 r \phi _1(r)= 0, \\
&&r \left(f_2'(r)+r \left(r^2+1\right) \phi _1'(r){}^2-2 r \phi _1(r){}^2\right)+f_2(r) = 0, \\ 
&&\delta _2'(r)+r \phi _1'(r){}^2= 0, 
\eeqa
which are solved by 
\beqa
&&\phi_1(r) = \frac{2 \tan ^{-1}(r)}{\pi  r}, \label{sol_solution_1}\\ 
&&f_2(r) = \frac{4 \tan ^{-1}(r) \left(\left(r+\frac{1}{r}\right) \tan ^{-1}(r)-1\right)}{\pi ^2 r}, \label{sol_solution_2}\\
&&\delta_2(r) = \frac{1}{2}-\frac{2 \left(\frac{1}{r^2+1}+\left(1-\frac{1}{r^2}\right) \tan ^{-1}(r)^2+\frac{2 \tan ^{-1}(r)}{r}\right)}{\pi ^2}. \label{sol_solution_3} 
\eeqa
The boundary conditions are as follows: 
\begin{itemize}
	\item We have demanded that $\phi_1$ is regular at the origin, and also that $r \phi_1(r)\rightarrow 1$ as $r\rightarrow \infty$ -this is needed in order to set $J$ as the actual source.
	\item We demand that $f_2$ is regular at the origin.  
	\item We set $\delta_2(\infty)= 0$ to work in the boundary time gauge.  
\end{itemize}

Once the solution is known, we can employ it to compute $\delta f_s$. For AdSJ, we have 
\beq
f_s = -\frac{m_0}{16 \pi G} - J \epsilon - \frac{1}{8 \pi G}\int_0^\infty dr \left(e^{-\delta(r)} -1\right). 
\eeq
From our solution,  at leading order 
\beqa
&& m_0= (6/\pi) J^2 , \\
&& \epsilon = - (2/\pi) J, \\
&&\int_0^\infty dr \left(e^{-\delta(r)} -1\right) = -(1/\pi)J^2. 
\eeqa
In our units ($4 \pi G = 1$), we have 
\beq
f_s = \delta f_s J^2 = \frac{1}{\pi} J^2
\eeq
or $\delta f_s = 1/\pi$. 

The final step is finding the first correction in $J$ to the Schwarzschild black hole of $r_h = 1$. Again, we linearize our fields as 
\beqa\label{pert_expansion}
&&\phi(r) = J \phi_1(r), \\
&&f(r)  = f_{0}(r) + J^2 f_2(r), \\
&&\delta(r) = J^2 \delta_2(r), 
\eeqa
where $f_0(r) = -\frac{m_0}{r}+r^2+1$. For the black hole we are interested in, $m_0 = 2$. The scalar field equation reduces to 
\beq
\frac{2 \left(\left(2 r^3+r-1\right) \phi _1'(r)+r^2 \phi _1(r)\right)}{r \left(r^3+r-2\right)}+\phi _1''(r) = 0, 
\eeq
and, as for the AdSJ case,  $f_2, \delta_2$ can be determined by a simple integration once $\phi_1$ is known. Unfortunately, we have been unable to solve this equation analytically. We have proceeded by integrating it numerically. The equation is linear; we fix the freedom in rescaling $\phi_1$ by demanding that $r \phi_1(r)\rightarrow 1$ as $r\rightarrow \infty$, which corresponds again to setting $J$ as the actual field theory source. In this case, the free energy is given by 
\beq
f_{BH}(1) = -\frac{m_0}{16 \pi G} - J \epsilon + \frac{1}{8 \pi G} - \frac{1}{8 \pi G}\int_{1}^{\infty} dr \left(e^{-\delta(r)} -1\right). 
\eeq
At $O(J^2)$, 
\beq
f_{BH}(1) = \delta f_{BH}(1) J^2 = -\frac{\delta m_0 J^2}{16 \pi G} - \delta \epsilon J^2 - \frac{1}{8 \pi G}\int_{1}^{\infty} dr \left(- \delta_2(r) J^2 \right).
\eeq
We have 
\beqa
&&\delta m_0 = \frac{r^4}{6} f_2^{(3)}\left(r\right)|_\infty = 1.967, \\
&&\delta \epsilon = -0.6555, \\
&&\int_{1}^{\infty} dr \delta_2(r) = 0.2349.  
\eeqa
Therefore, 
\beq
\delta f_{BH}(\infty) = 0.2813. 
\eeq
From eq. \eqref{x_c_correction}, we find that the phase transition takes place at a horizon radius such that 
\beq
\delta r_c = - 0.9296. 
\eeq
Therefore, the radius of the first thermodynamically dominant black hole decreases with respect to its position at $J = 0$. Finally, the temperature of this black hole, i.e., the Hawking-Page transition temperature, changes as
\beq
T_{HP}(J) = T_{HP}(0) + J^2 \delta T_{HP}= \frac{1}{\pi} + J^2 \frac{2 \delta r_c - 4 \delta_2 (1) + f_2'(1)}{4 \pi} = \frac{1}{\pi} + 0.003011344 J^2. \label{THP_corr_1}
\eeq
The Hawking-Page temperature increases with increasing coupling. To obtain the expression above, we have expanded the formula for the black hole temperature to $O(J^2)$, 
\beq
T_{BH} = \frac{1}{4 \pi} f'(r_h) e^{- \delta(r_h)} = \frac{\frac{1}{r_h} + 3 r_h}{4 \pi} + \frac{f_2'(r_h) - \delta_2(r_h)(\frac{1}{r_h} + 3 r_h)}{4 \pi} J^2 
\eeq
and evaluated it at the phase transition point, $r_h = 1 + \delta r_c J^2$, keeping only terms up to $O(J^2)$. 

\begin{figure}[h!]
	\begin{center}
		\includegraphics[width=12cm]{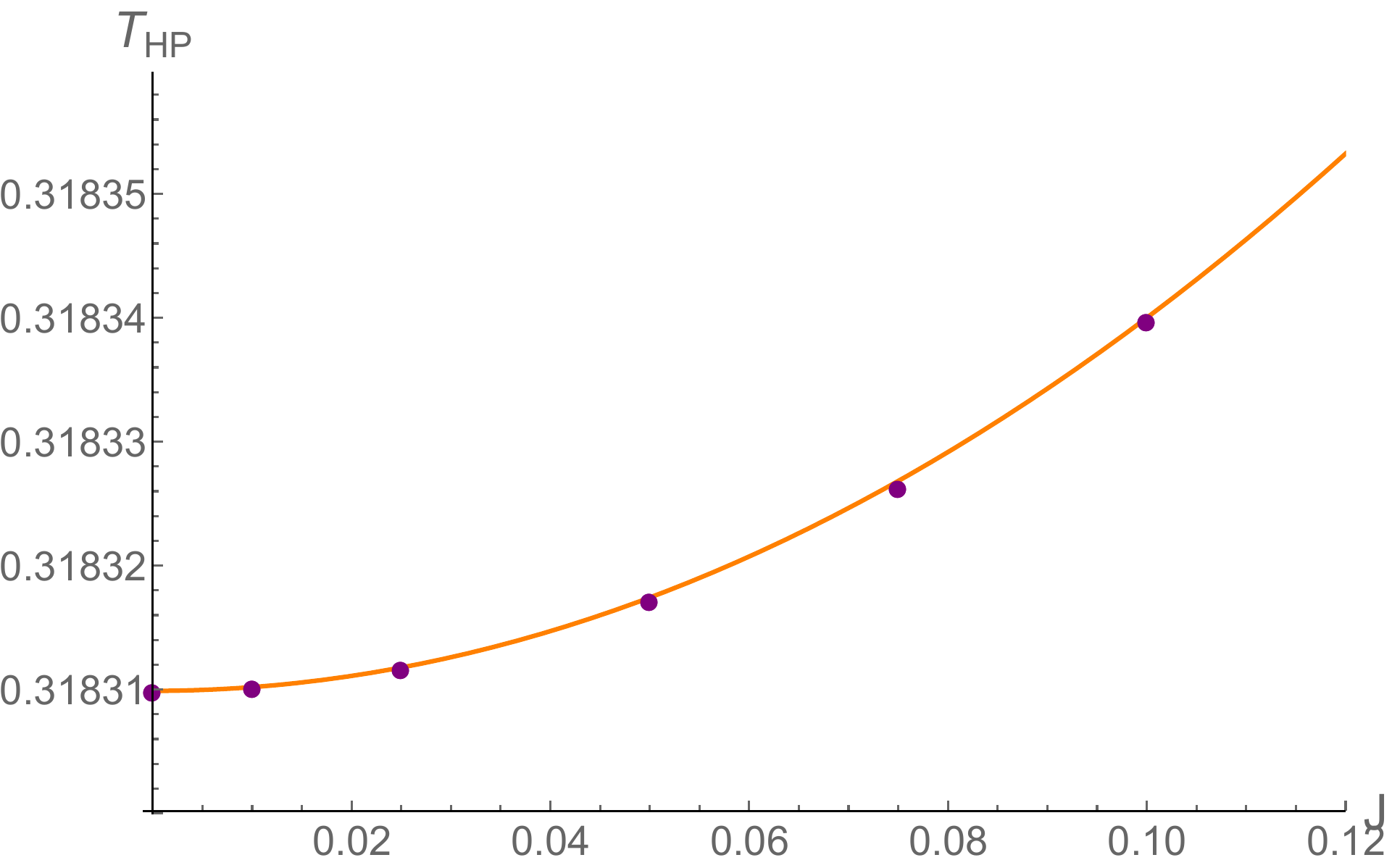}
		\caption{\small Perturbative prediction for $T_{HP}$ -solid orange- and actual values coming from a full fledged numerical computation -purple dots-. The agreement is extremely good.}
		\label{fig:T}
	\end{center}
\end{figure}

Numerical results are compared to this perturbative formula in Fig.\,\ref{fig:T}.

\subsection{Second method}
\label{THP_pert_2}

Let me introduce, without writing the explicit proof, the following identities\footnote{I am employing our $4 \pi G = L = 1$ units.}
\beq
f = \frac{1}{2}m_0 + J \epsilon - T s, \label{smarr}
\eeq
\beq
T ds = \frac{1}{2} dm_0 + J d\epsilon + 2 \epsilon dJ. \label{first_law}
\eeq
Taking a variation of \eqref{smarr}, and with the help of \eqref{first_law}, 
\beq
df = \frac{1}{2} dm_0 + J d\epsilon + \epsilon dJ - T ds - s dT =  - s dT - \epsilon dJ. 
\eeq
Upon an infinitesimal change in $J$, $J \rightarrow J + dJ$, the phase transition point changes as 
\beq
f_s + df_s = f_{BH} + d f_{BH}.  
\eeq
Taking into account that at the phase transition point $f_s = f_{BH}$, we obtain 
\beq
0 = df_s - d f_{BH} = \left(-\epsilon_s + s_{BH} \frac{dT_{HP}}{dJ} + \epsilon_{BH}\right) dJ. 
\eeq
The above relation allows us to determine $T_{HP}'(J)$, 
\beq
T_{HP}'(J) = \frac{1}{s_{BH}}\left(\epsilon_s - \epsilon_{BH} \right). \label{master}
\eeq
The quantities $s_{BH}, \epsilon_{BH}$ and $\epsilon_s$ are evaluated at the Hawking-Page temperature at coupling $J$. 

We know that, at first nontrivial order in $J$, $\epsilon_s = -2/\pi J$ and $\epsilon_{BH} = -0.6555...J$. The Hawking-Page transition in the absence of a source happens at $x_c^0 = \pi/4$, and therefore $s_{BH} = 1/(4 G) = \pi$ in our units. Then, 
\beq
T_{HP}'(J) = \frac{1}{\pi} \left(-2/\pi + 0.6555...\right) J.    
\eeq
Integrating (with the boundary condition that $T_{HP}(0) = 1/\pi$), 
\beq
T_{HP}(J) = \frac{1}{\pi} + 0.00301128 J^2 + O(J^4). \label{THP_corr_2}
\eeq
We clearly see that the corrections determined by both methods agree very well; the relative difference between \eqref{THP_corr_1} and \eqref{THP_corr_2} is $< 0.01\%$, and most likely due to accumulated numerical error. 
\\\\

\section{Same story Different quantization: Alternative quantization}
\label{sec:alternative}

In the alternative quantization, the physical roles played by $J$ and $\epsilon$ are interchanged: now $\epsilon$ corresponds to the source of a $\Delta = 1$ scalar operator, while $J$ corresponds to its vacuum expectation value. Therefore, in this section we are interested in the dependence of the Hawking-Page temperature with $\epsilon$.

\subsection{Perturbative computation of the Hawking-Page temperature}

Once expressions \eqref{f_x_alt} and \eqref{smarr_alt} are known, we can compute $T_{HP}$ in the alternative quantization along the lines of subsection \ref{THP_pert_1} or \ref{THP_pert_2}.\footnote{Note that the First Law \eqref{first_law} is unchanged, since it does not depend on whether we identify $J$ as the source and $\epsilon$ as the vev or vice versa.} For concreteness, we will employ the Clausius-Clapeyron version of the derivation. 

From \eqref{smarr_alt}, we have that 
\beq
df = \frac{1}{2}dm_0 + 2 J d\epsilon + 2 \epsilon dJ - T ds - s dT. 
\eeq
Using the First Law \eqref{first_law} in the expression above, we get 
\beq
df = - s dT + J d\epsilon. \label{df_alt}
\eeq
Moving along the phase transition curve, we have again that $f_s + \frac{df_s}{d\epsilon} d\epsilon = f_{BH} + \frac{df_{BH}}{d\epsilon} d\epsilon$. Since $f_s = f_{BH}$ along the curve, 
\eqref{df_alt} implies that 
\beq
\frac{d T_{HP}}{d\epsilon} = \frac{1}{s_{BH}}(J_{BH} - J_s), \label{master_alt}
\eeq 
which is equivalent to \eqref{master} with the substitutions $J \to \epsilon$, $\epsilon \to - J$. 

Again, equation \eqref{master_alt} is an exact statement, valid at any source $\epsilon$. To employ it to compute the first perturbative correction to $T_{HP}$, we must determine how $J$ depends on $\epsilon$ for $\epsilon \ll 1$, both for the AdSJ solution and the hairy black hole of horizon radius $x_h = \pi/4$. 

A strategy analogous to the one employed in subsection \ref{THP_pert_1} tells us that 
\beq
J_s = -\frac{\pi}{2}\epsilon + O(\epsilon^2),~~~~~~J_{BH} = -1.5255 \epsilon + O(\epsilon^2). 
\eeq
Recaling that at $\epsilon = 0$, $s_{BH} = 1/(4 G) = \pi$
\beq
T_{HP}'(\epsilon) = \frac{1}{\pi} (-1.5255 + \pi/2) \epsilon = 0.0144319 \epsilon. 
\eeq
Taking into account that $T_{HP}(\epsilon = 0) = 1/\pi$, we finally have that 
\beq
T_{HP}(\epsilon) = \frac{1}{\pi} +  0.00721596 \epsilon^2 + O(\epsilon^4). \label{T_HP_alt}
\eeq

\begin{figure}[h!]
	\begin{center}
		\includegraphics[width=12cm]{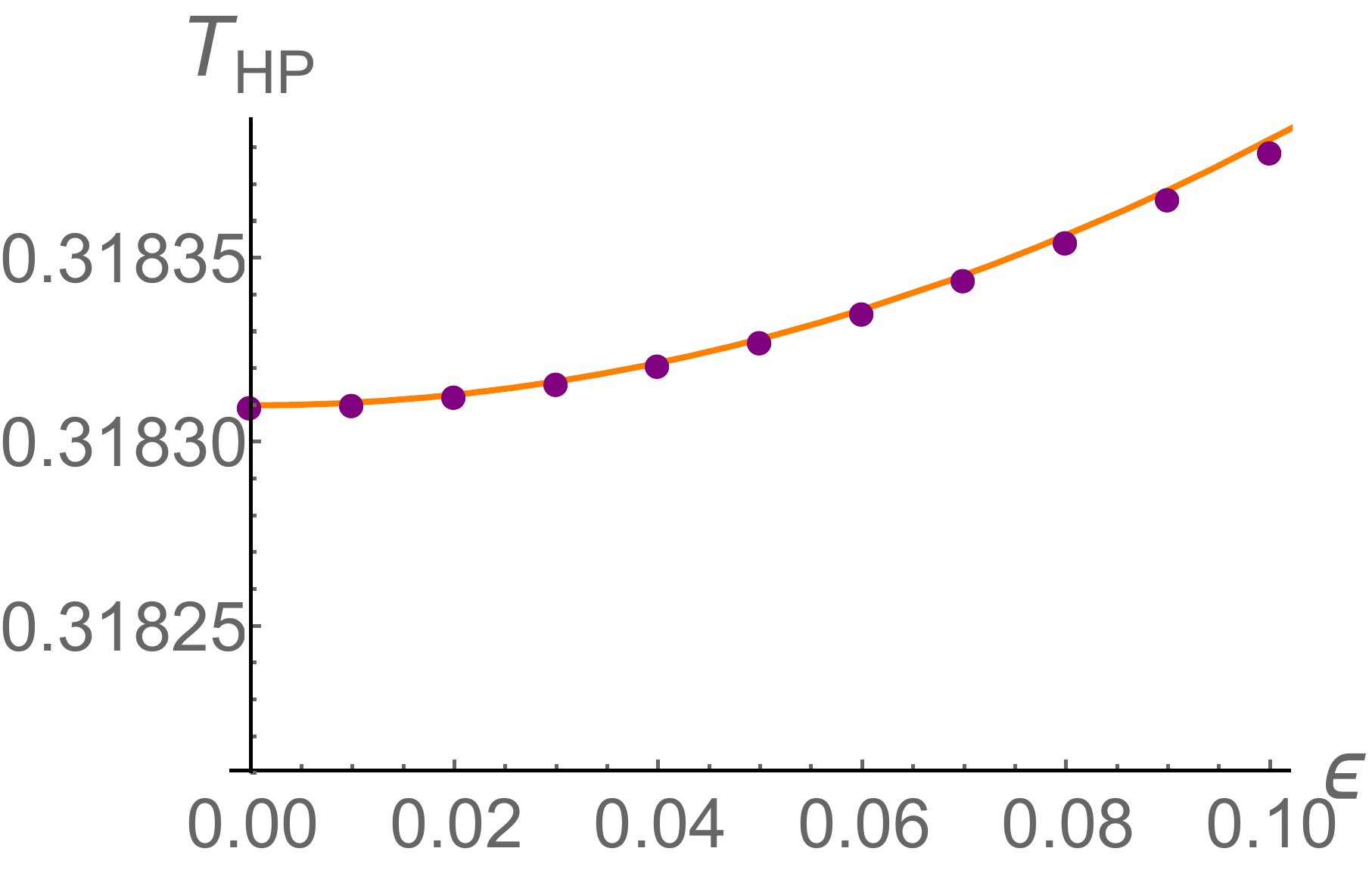}
		\caption{\small Perturbative prediction for $T_{HP}$ -solid orange- and actual values coming from a full fledged numerical computation -purple dots-. The agreement is extremely good.}
		\label{fig:T_alt}
	\end{center}
\end{figure}

As for the standard quantization of the scalar field, the perturbative prediction is that the Hawking-Page temperature increases with the relevant coupling. In Fig.\,\ref{fig:T_alt} we contrast this  
expectation against exact numerical results in the small coupling regime, finding excellent agreement.  

\subsection{Fully nonlinear results}

\begin{figure}[h!]
	\begin{center}
		\includegraphics[width=12cm]{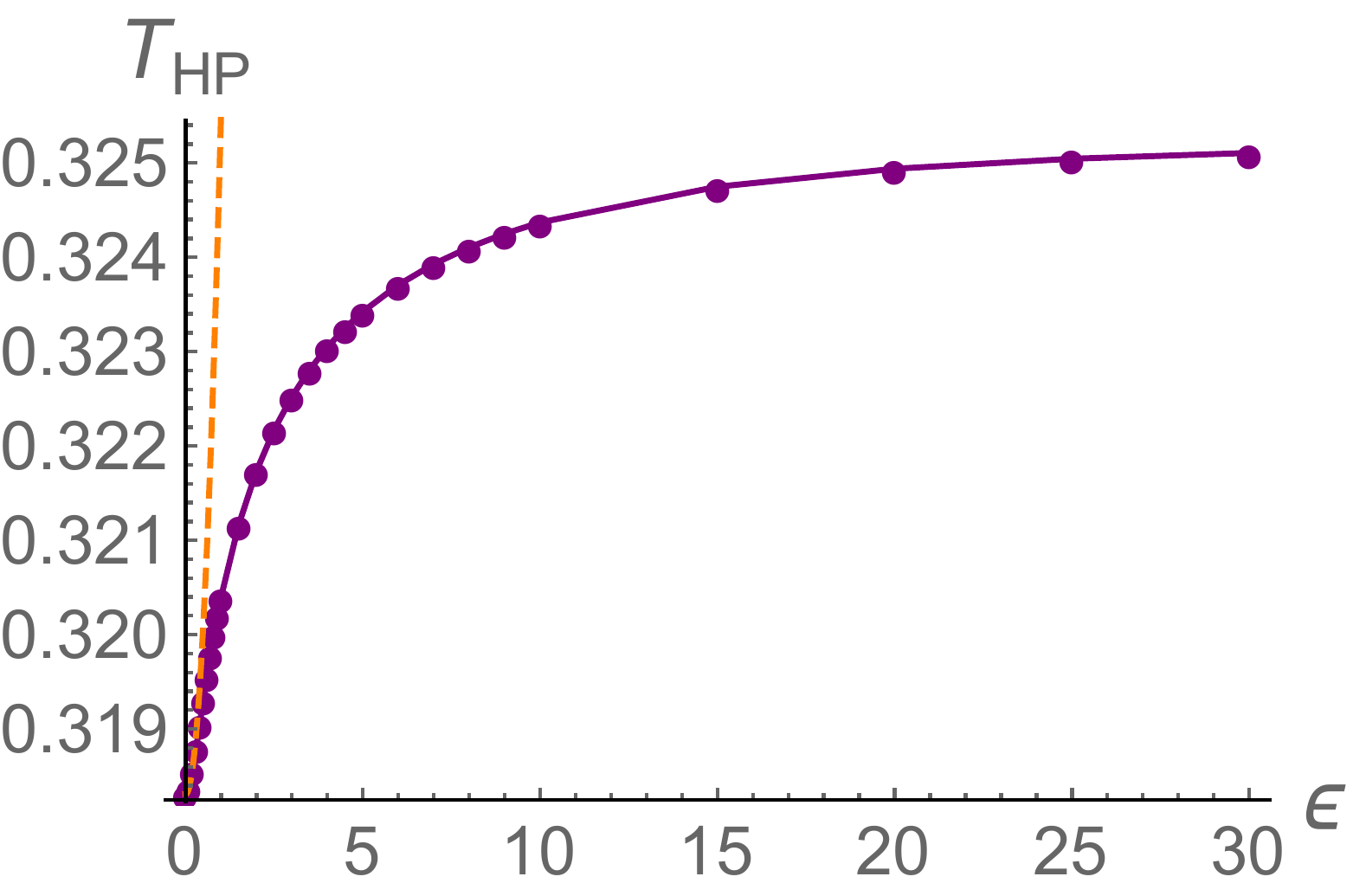}
		\caption{\small Phase transition temperature $T_{HP}$ vs relevant coupling $\epsilon$ (purple dots). The leading order perturbative prediction for $T_{HP}$ (eqn. \eqref{THP_corr_1}) corresponds to the dashed orange line.}
		\label{fig:T_fullynonlinear_alt}
	\end{center}
\end{figure}

In this case, and in contrast to what happened in the standard quantization, our numerical results\footnote{We have employed the same pseudospectral method as before; going from the standard to the alternative quantization just amounts to a trivial change in the scalar field boundary conditions.} show that the Hawking-Page temperature is monotonic, saturating to a finite value as $\epsilon \to \infty$ (see Fig.\,\ref{fig:T_fullynonlinear_alt}). Again, this is compatible with the absence of a planar limit for the Hawking-Page transition.

\begin{figure}[t!]
	\begin{center}
		\includegraphics[width=11cm]{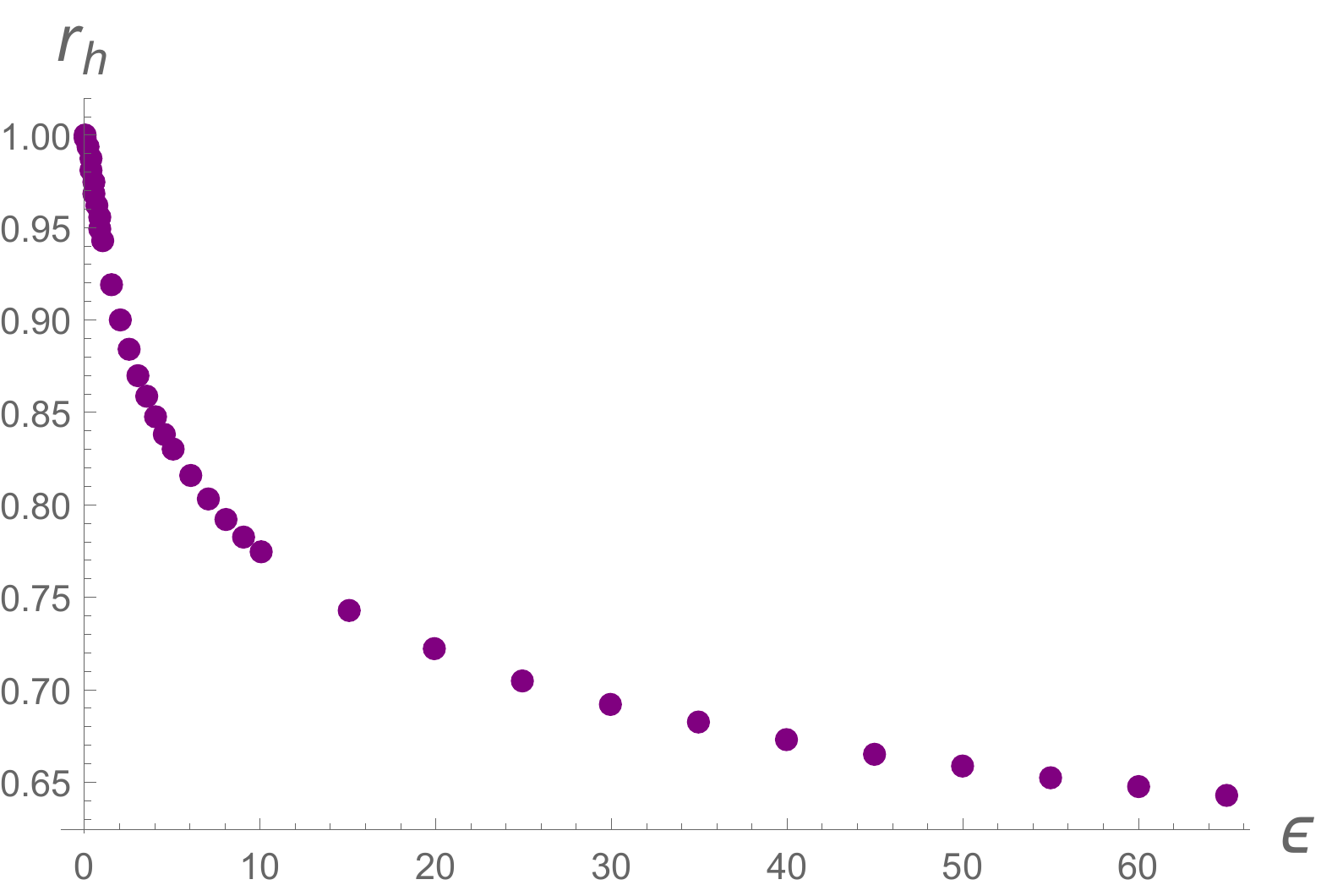}
		\caption{\small Radius of the first thermodynamically dominant black hole in terms of the coupling.}
		\label{fig:radius_alt}
	\end{center}
\end{figure}

As in the standard quantization case, the Hawking-Page temperature is mildly affected by the increasing relevant coupling (for instance, for the last coupling we have considered, $\epsilon = \epsilon_{max} = 65$, we have that $(T_{HP}(\epsilon_{max}) - T_{HP}(0))/T_{HP}(0) \approx 2.15 \%$). As before, the entropy released at the phase transition also shows a marked  
decrease (see Fig.\,\ref{fig:radius_alt}). 

\subsection{Normal modes of the AdS$\epsilon$ solution}

In order to determine the normal modes of the AdS$\epsilon$ solution, we need to solve equation \eqref{eq_normal_mode} with the following boundary conditions for $\phi_1$ 
\begin{itemize}
	\item Origin regularity, i.e., $\phi_1(0) < \infty$. 
	\item Normalizability. The source $\epsilon$ that supports the AdSJ is kept invariant by the perturbation, i.e., $\phi_1''(\pi/2) = 0$. 
\end{itemize}

\begin{figure}[h!]
	\begin{center}
		\includegraphics[width=16cm]{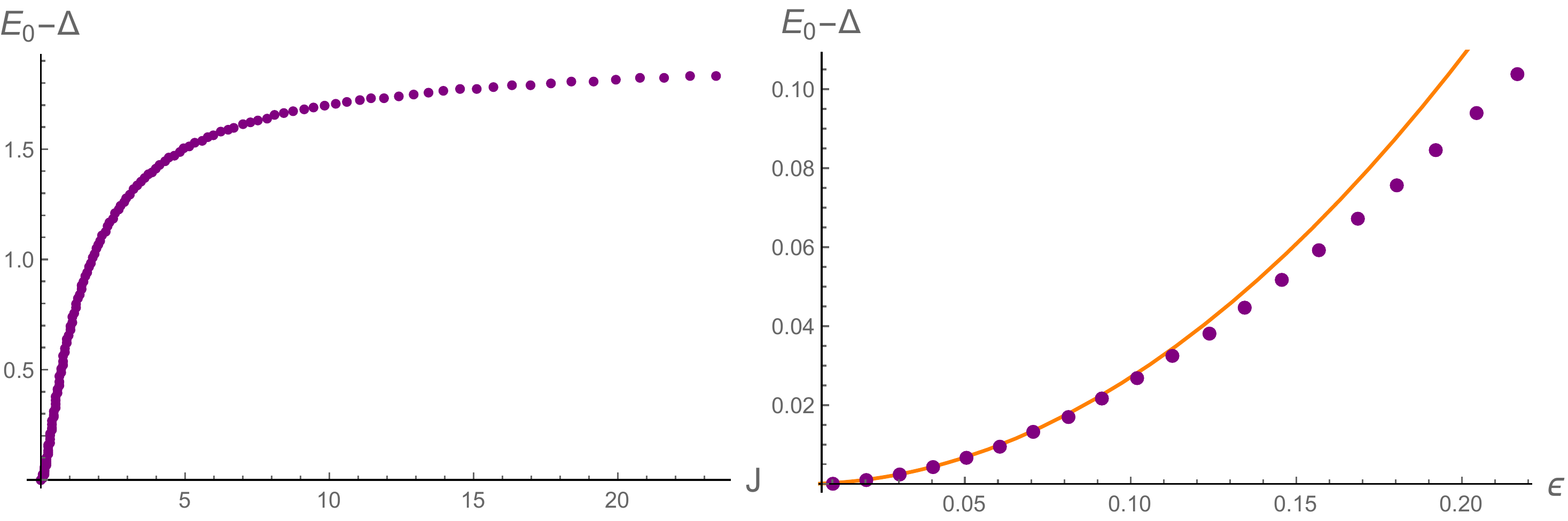}
		\caption{\small Left: Difference between the lowest eigenfrequency of the AdSJ, $E_0$, and its value on the AdS$_4$ vacuum, $\omega_0 = \Delta = 1$, as a function of the source $\epsilon$ supporting the AdSJ background. Right: as in the left plot, but for smaller $\epsilon$. The solid orange line is the leading order dependence of $E_0$ on $\epsilon$ predicted by perturbation theory.}
		\label{fig:eigenfrequency_alt}
	\end{center}
\end{figure}

Using the same pseudospectral code as before, we get the results displayed in Fig.\,\ref{fig:eigenfrequency_alt}a. Unlike in the standard quantization case, here the behavior of the lowest eigenfrequency of the AdSJ is correlated with the behavior of the Hawking-Page temperature. 

The first correction to the lowest eigenfrequency can be computed again in an expansion in $\epsilon$. The result is 
\beq
\omega_0 = 1 + \left(\frac{3 \pi^2}{8} - 1 \right) \epsilon^2 + O(\epsilon^4)
\eeq
and is compared to the exact numerical values in Fig.\,\ref{fig:eigenfrequency_alt}b, finding very good agreement. 

\section{Conclusion}
In this work we discussed the how a compact boundary cuts down the holographic RG flow leading to an incomplete RG flow. We further discussed the implication of the incomplete RG flow for various physically relevant issues including black hole formation and deconfinement transition. We believe our results to be true in other dimensions and relevant scalar operators with different dimensions. It would be interesting to investigate physics of double trace deformations \cite{2001hep.th12258W} in finite temperature and with a compact boundary.  Another natural extension of our work may be in the case of the magnetic branes \cite{D_Hoker_2009} and holographic lattices \cite{Horowitz_2012}. Instead of a theory with compact boundary, if we consider a confining theory, like say AdS solition with boundary $S^2\times S^1 \times S^1$, then also we hope to  phenomena similar to incomplete RG flow   \cite{toappear}. We noted how turning on a relevant scalar operator in the boundary introduces negative energy in the bulk; virtually increasing the effective cosmological constant. We may speculate whether this negative energy may be used to sustain an worm hole \cite{Bronnikov_2018} and whether there are alternative ways to sustain a traversable wormhole \cite{maldacena2020traversable}.
\section{Acknowledgment}
We thank Alexandre Serantes Rubianes for discussions and collaboration at various stages of this project. PB thanks  Jyotirmoy Bhattacharya, Vishnu Jejjala for discussions. PB also thanks  Robert De Mello Koch  and other organizers of MITP including workshop in Durban 2020, where this work was presented.  
\appendix

\section{An alternative co-ordinate system}\label{app:tanx}
Another alternative co-ordinate system to use $(t,r,\theta,\phi)\rightarrow(t,x,\theta,\phi)$ with $\tan(x)=r$. This co-oridante has the advantage of a finite domain  $x\in [0,\frac{\pi}{2}]$. In our numerical analysis we often use this coordinate because a finite domain often facilitates the numerics and makes plots more revealing. We use following spherically symmetric metric ansatz: 
\beq
ds^2 = \sec^2 x (- f e^{-2 \delta} dt^2 + f^{-1}dx^2 + \sin^2 x d\Omega_{d-1}^2), \label{ds3}
\eeq
where $\delta$, $f$ and the scalar field $\phi$ are functions of $t$ and $x$. $d\Omega_{d-1}^2$ is the metric on $d-1$ sphere.  If $f=1,\delta=0$, the above ansatz becomes a global AdS$^{d+1}$ space. We note that definition of the blackening factor in \eqref{ds3} is slightly different than \eqref{ds2}.

For the static and symmetric solutions relevant for the computation of the phase diagram, the equations reduce to 
\beqa\label{eqns}
&&\delta '(x)+\sin (x) \cos (x) \phi '(x)^2 = 0, \nonumber  \\
&&f'(x)+f(x) \sin (x) \cos (x) \phi '(x)^2+(f(x)-1) \tan (x) \left(\csc ^2(x)+2\right)-2 \phi (x)^2 \tan (x)=0, \nonumber\\
&&\phi ''(x)  + \left(\frac{f'(x)}{f(x)}-\delta '(x)+2 \csc (x) \sec (x)\right)\phi '(x) +\frac{2 \sec ^2(x)}{f(x)}\phi (x)= 0. 
\eeqa

\section{Holographic renormalization}\label{app:renorm}
\label{app:global}

\subsection{Holographic renormalization: standard quantization}

Let us consider first the variation of the scalar field action. After integrating by parts and employing the equations of motion, we are left with 
\beq
\delta S_\phi = - \int_{r=R} dt d\theta d\varphi \sqrt{-h} n^r \partial_r \phi \delta \phi, \label{var_scalar_action}
\eeq
where $n^\mu$ is the outward unit normal to a constant $r=R$ hypersurface, 
\beq
n^\mu = (0,\sqrt{F(r)},0,0) \label{unit_normal}
\eeq
and $\sqrt{-h}=r^2 e^{-\delta(r)} \sqrt{F(r)} \sin \theta$ is the determinant of the induced metric on this hypersurface,
\beq
dh^2 = - F(r) e^{-2 \delta(r)} dt^2 + r^2 d\Omega^2. \label{dh2}
\eeq
For scalar field fluctuations $\delta \phi$ such that $\delta J = 0$, 
\beq
\delta \phi = \frac{\delta \epsilon}{r^2} + \ldots,
\eeq
the variation \eqref{var_scalar_action} reduces to 
\beq
\delta S_\phi = \int_{r=R} dt d^2 \Omega (J \delta\epsilon) + O(1/R)
\eeq
where $d^2 \Omega = \sin \theta d\theta d\varphi$, and the asymptotic expansions of $F, \delta$ and $\phi$ have been employed. We observe that $\delta S_\phi$ does not vanish on the static geometries with finite source we are interested in. To correct this, we add the following countertem
\beq
S_{ct, \phi} = -\frac{1}{2}\int_{r=R} dt d\theta d\varphi \sqrt{-h} \phi^2. \label{S_ct_phi}
\eeq
In is immediate to check that, with this modification, the scalar field action is now stationary on solutions with finite $J$,
\beq
\delta (S_{\phi} + S_{ct,\phi}) = 0, 
\eeq
as long as we restrict ourselves to scalar variations $\delta \phi$ that don't change the source. 

Let us discuss now the holographic renormalization of the full action \eqref{action}. Owing to the spherical symmetry of the background spacetime, we can write the total Lagragian density 
\beq
\mathcal L = \frac{1}{16 \pi G} (R + 6) - \left(\frac{1}{2}(\partial \phi)^2 + V(\phi) \right)\label{lagrangian}, 
\eeq
in terms of the Einstein tensor as
\beq
\mathcal L = - \frac{1}{16 \pi G} \left(G^t_t + G^r_r \right). 
\eeq
Therefore, the bulk action (forgetting about boundary terms) is given by 
\beq
S_{bulk} = \int d^4 x \sqrt{-g} \mathcal L = \int dt dr d^2 \Omega \left(-\frac{e^{-\delta(r)}(-1+F(r) + r F'(r) - r F(r) \delta'(r))}{8 \pi G} \right). 
\eeq
The integrand in the expression above can be rewritten as a total radial derivative plus a non-exact piece, 
\beq
\sqrt{-g}\mathcal L = -\frac{1}{8 \pi G} \partial_r (r e^{-\delta(r)}F(r)) + \frac{e^{-\delta(r)}}{8 \pi G}.\label{integrand} 
\eeq

\noindent Upon radial integration, the term inside the total radial derivative vanishes in the infrared end of the geometry, $r = r_{IR}$.\footnote{For a black hole horizon $r_{IR} = r_h$ and $F(r_h) = 0$; for a soliton, $r_{IR} = 0$ and the remaining functions are finite there by regularity.} Therefore, only the upper boundary of the integration region contributes. From the asymptotic expansions, it follows that this contribution is divergent
\beq
- \frac{r e^{-\delta(r)} F(r)}{8 \pi G} =-\frac{r^3}{8 \pi  G}+r \left(-\frac{1}{8 \pi  G}-\frac{J^2}{4}\right)+\left(\frac{2 \epsilon  J}{3}+\frac{m_0}{8 \pi  G}\right)+ O(r^{-1}). 
\eeq
The radial integrand of the non-exact piece is also $\sim r$ asymptotically, since $\delta(r) \to 0$ as $r \to \infty$. In order to better handle this divergence, we add and substract $1$ to the integrand. We are left with
\beq
\int d^4 x \frac{e^{-\delta(r)}-1+1}{8 \pi G} = \int dt d^2 \Omega \int_{r= r_{IR}}^{r= R} \frac{e^{-\delta(r)}-1}{8 \pi G} + \int dt d^2 \Omega\frac{R - r_{IR}}{8 \pi G}. 
\eeq 
The linearly divergent piece adds to the overall divergence of the bulk action, which is given by the integral of 
\beq
-\frac{r^3}{8 \pi G} - \frac{J^2 r}{4} + \left(\frac{m_0}{8 \pi G} + \frac{2 J \epsilon}{3} \right) + O(r^{-1}) 
\eeq
over the $r = R$ hypersurface. 

This completes the discussion of the bulk action. Now, we have to take into account that, in order to have a well-posed variational problem for the metric, we have to add the Gibbons-Hawking term, 
\beq\label{GHterm}
S_{GH} = \frac{1}{8 \pi G} \int_{r=R} dt d\theta d\varphi \sqrt{-h} K, 
\eeq
where $K$ is the trace of the extrinsic curvature of  $r = R$ hypersurface, and $\sqrt{-h}$ is given right before \eqref{dh2}. In our case, $K$ can be expressed as 
\beq
K = \nabla_\mu n^\mu = \frac{1}{\sqrt{-g}} \partial_\mu (\sqrt{-g} n^\mu), 
\eeq
where $n^\mu$ is given by \eqref{unit_normal}. In the end, the asymptotic series expansions tell us that the integrand of the Gibbons-Hawking term is divergent, 
\beq
\frac{\csc \theta}{8 \pi G}\sqrt{-h} K = \frac{3 r^3}{8 \pi  G} +  \left(\frac{1}{4 \pi  G}+\frac{3 J^2}{4}\right)r - \frac{3 m_0}{16 \pi  G}+ O(r^{-1}).  
\eeq
The combined divergence of $S_{bulk}$ and $S_{GH}$, 
\beq
(S_{bulk} + S_{GH})_{div} = \int_{r=R} dt d^2 \Omega \left(\frac{r^3}{4 \pi  G}+r \left(\frac{1}{4 \pi  G}+\frac{J^2}{2}\right)+\left(-\frac{m_0}{16 \pi  G}+\frac{2 J \epsilon}{3}\right)+O(r^{-1}) \right),  \label{bulk_div}
\eeq
has to be cancelled by the introduction of the appropriate counterterms. Besides the scalar counterterm $S_{ct,\phi}$, these are 
\beqa
&&S_{ct,1} = -\frac{1}{4 \pi G} \int_{r=R} dt d\theta d\varphi \sqrt{-h}, \\
&&S_{ct,2} = -\frac{1}{16 \pi G} \int_{r=R} dt d\theta d\varphi \sqrt{-h} R(h), 
\eeqa
where $R(h)$ is the Ricci scalar of the induced metric on the constant $R$ hypersurface. It is immediate to check that, with these modifications, the total action 
\beq
S = S_{bulk} + S_{GH} + S_{ct,1} + S_{ct,2} + S_{ct, \phi}
\eeq
is finite: each counterterm we had introduced has the following large $r$ expansion, 
\beqa
&&S_{ct,1} = -\frac{1}{4 \pi G} \int_{r= R} dt d^2 \Omega \left(r^3+\frac{r}{2}-\frac{m_0}{2}-\frac{16}{3} \pi  G J \epsilon + O(r^{-1})\right), \\
&&S_{ct,2} = - \frac{1}{16 \pi G} \int_{r=R} dt d^2 \Omega \left( 2 r + O(r^{-1})\right), \\
&&S_{ct,\phi} = - \frac{1}{2} \int_{r=R} dt d^2 \Omega \left(J^2 r + 2 J \epsilon+ O(r^{-1})\right). 
\eeqa
and  cancels the $O(r^3)$, $O(r)$ divergences of \eqref{bulk_div}, leaving only 
\beq
\int_{r=R} dt d^2 \Omega \left(\frac{m_0}{16 \pi G} + J \epsilon \right)
\eeq
as a finite contribution. 
\\\\
\noindent Finally, the holographically renormalized action after taking the UV cutoff $R \to \infty$ is 
\beq
\frac{S}{\Delta t \textrm{Vol}(S^2)} = \frac{m_0}{16 \pi G} + J \epsilon  - \frac{r_{IR}}{8 \pi G} + \int_{r_{IR}}^\infty dr \frac{e^{-\delta(r)}-1}{8 \pi G}, \label{S_lorentzian}
\eeq
where we have employed the fact that the geometry is time-independent. $\Delta t$ corresponds to the the range of integration along the coordinate $t$. Equation \eqref{S_lorentzian} is the on-shell action in Lorentzian signature. In order to analyze the thermodynamics of a given saddle, we need to perform an analytical continuation to Euclidean signature, $t \to -i \tau$, and identify the Euclidean time as $\tau \sim \tau + \beta$, where $\beta = 1/T$ is the length of the thermal cycle. This procedure just amounts to a sign flip of the Lorentzian action \eqref{S_lorentzian}. The free energy density is $f$ related to the Euclidean on-shell action as,  
\beq
S_E = \beta \textrm{Vol}(S^2) f. 
\eeq
The final result, which also appears in equation \eqref{f_x}in the main text, is  
\beq
f = -\frac{m_0}{16 \pi G} - J \epsilon  + \frac{r_{IR}}{8 \pi G} + \int_{r_{IR}}^\infty dr \frac{1-e^{-\delta(r)}}{8 \pi G}. \nonumber
\eeq

As first consistency check, let us spell out the computation the Hawking-Page transition temperature at zero coupling. The thermal AdS metric is given by $f(r) = 1, \delta(r) = \phi(r) = 0$, and therefore $f_{EAdS} = 0$. On the other hand, for a Schwarzschild black hole, we have that $\delta(r) = \phi(r) = 0$ and 
\beq
f = 1 - \frac{m_0}{r(1+r^2)}. 
\eeq
In terms of the horizon radius $r_h$, $m_0 = r_h(1+r_h^2)$. The free energy density is 
\beq
f_{SAdS} = \frac{1}{16 \pi G}r_h (1- r_h^2). 
\eeq
For $r_h > 1$, $f_{SAdS} < 0$ and the black hole solution is thermodynamically dominant. The deconfinement transition takes place between a black hole of horizon radius $r_h = 1$ and thermal AdS, which corresponds to a temperature $T = 1/ \pi$, as expected.

\subsection{Holographic renormalization: alternative quantization}

In this case, we are focusing on scalar field variations of the form
\beq 
\delta \phi = \frac{\delta J}{r} + O(r^{-3}).
\eeq
For these variations, the scalar field action variation \eqref{var_scalar_action} is given by 
\beq
\delta S_\phi = - \int_{r = R} dt d^2\Omega \left(J \delta J r + 2 \epsilon \delta J + ... \right). 
\eeq
In order to render $\delta S_\phi$ finite, we must add again the counterterm \eqref{S_ct_phi}. However, in this case, the addition of this counterterm does not render automatically the scalar field variational problem well-posed. We have that
\beq
\delta \left(S_\phi + S_{ct,\phi} \right) = \int_{r=R} dt d^2 \Omega \left(\epsilon \delta J + ... \right)
\eeq
In order to have a stationary action, we add yet another new counterterm
\beq
S'_{ct,\phi} = - \frac{1}{R^2} \int_{r = R} dt d^2 \Omega \sqrt{- h} \epsilon \phi. 
\eeq
This new countertem has the property that
\beq
\delta S'_{ct,\phi} = \int_{r = R} dt d^2 \Omega \sqrt{- h}\left( -\epsilon \delta J + ...\right), 
\eeq
and therefore $\delta(S_{\phi} + S_{ct,\phi} + S'_{ct,\phi}) = 0$. On shell, 
\beq
S'_{ct,\phi} = \int_{r=R} dt d^2 \Omega (- J \epsilon). 
\eeq
So far, we have been working in Lorentzian signature. Upon analytical continuation to Euclidean signature, the action is going to pick up an extra minus sign. This entails that the presence of the new $S'_{ct,\phi}$ counterterm is shifting the free energy density as $f \to f + J \epsilon$. Expression \eqref{f_x} for the free energy density thus becomes 
\beq
f = -\frac{m_0}{16 \pi G} + \frac{\tan x_{IR}}{8 \pi G} - \int_{x_{IR}}^\frac{\pi}{2} dx \sec^2 x \frac{e^{-\delta(x)}-1}{8 \pi G}. \label{f_x_alt}
\eeq
\subsection{The planar limit of the AdSJ branch in the global case}
\label{app:planar-limit}

\subsubsection{The planar limit: definition \& example}

The asymptotic behavior of our fields is (in standard Schwarzschild coordinates): 
\beq
\begin{split}
	ds^2 =&- \left(r^2 + 1 - \frac{m_0 + 32/3 \pi G J \epsilon}{r} + ...\right) dt^2 + \frac{dr^2}{r^2 + 1 + 4 \pi G J^2 - \frac{m_0}{r}+...} +  \\
	&r^2 d\theta^2 + r^2 \sin^2 \theta d\varphi^2,
\end{split}
\eeq
\beq
\phi = \frac{J}{r} + \frac{\epsilon}{r^2} + ...
\eeq
Consider the following coordinate change 
\beq
r = \lambda \hat r,~~~~~~t = \hat t/\lambda,~~~~~~\theta = \hat \theta/\lambda, 
\eeq
and redefinitions, 
\beq
J = \hat J \lambda,~~~~~~\epsilon = \hat \epsilon \lambda^2,~~~~~~m_0 = \hat m_0 \lambda^3. 
\eeq
We have that 
\beq
\begin{split}
	ds^2 =&- \left(\hat r^2 + 1/\lambda^2 - \frac{\hat m_0 + 32/3 \pi G \hat J \hat \epsilon}{\hat r} + ...\right) d\hat t^2 + \frac{d\hat r^2}{\hat r^2 + 1/\lambda^2 + 4 \pi G \hat J^2 - \frac{\hat m_0}{\hat r}+...} +  \\
	& \hat r^2 d \hat \theta^2 + \lambda^2 \hat r^2 \sin^2 (\hat \theta/\lambda) d\varphi^2,
\end{split}
\eeq
\beq
\phi = \frac{\hat J}{\hat r} + \frac{\hat \epsilon}{\hat r^2} + ...
\eeq
So far, this is just a coordinate change. However, if we take the singular $\lambda \to \infty$ limit and, simultaneously, move along a one-parameter family of solutions in such a way that the hatted quantities remain finite, we obtain a planar geometry characterized by these hatted quantities. The fact that the limiting geometry is planar is easily established by noting that, as $\lambda \to \infty$, the $S^2$ part of the metric reduces to a plane in polar coordinates 
\beq
\hat r^2 d \hat \theta^2 + \lambda^2 \hat r^2 \sin^2 (\hat \theta/\lambda) d\varphi^2 \longrightarrow  \hat r^2 (d \hat \theta^2 +  \hat \theta^2 d\varphi^2), 
\eeq
with $\hat \theta$ playing the role of the radial coordinate on the plane. In particular, if we take $\lambda = J$, $\hat J = 1$, we observe that our one-parameter family of geometries has a nontrivial planar limit as long as the ratios 
\beq
\hat \epsilon = \frac{\epsilon}{J^2},~~~~~~\hat m_0 = \frac{m_0}{J^3} \label{ratios}
\eeq
remain finite as $J \to \infty$. 

\begin{figure}[h!]
	\begin{center}
		\includegraphics[width=16cm]{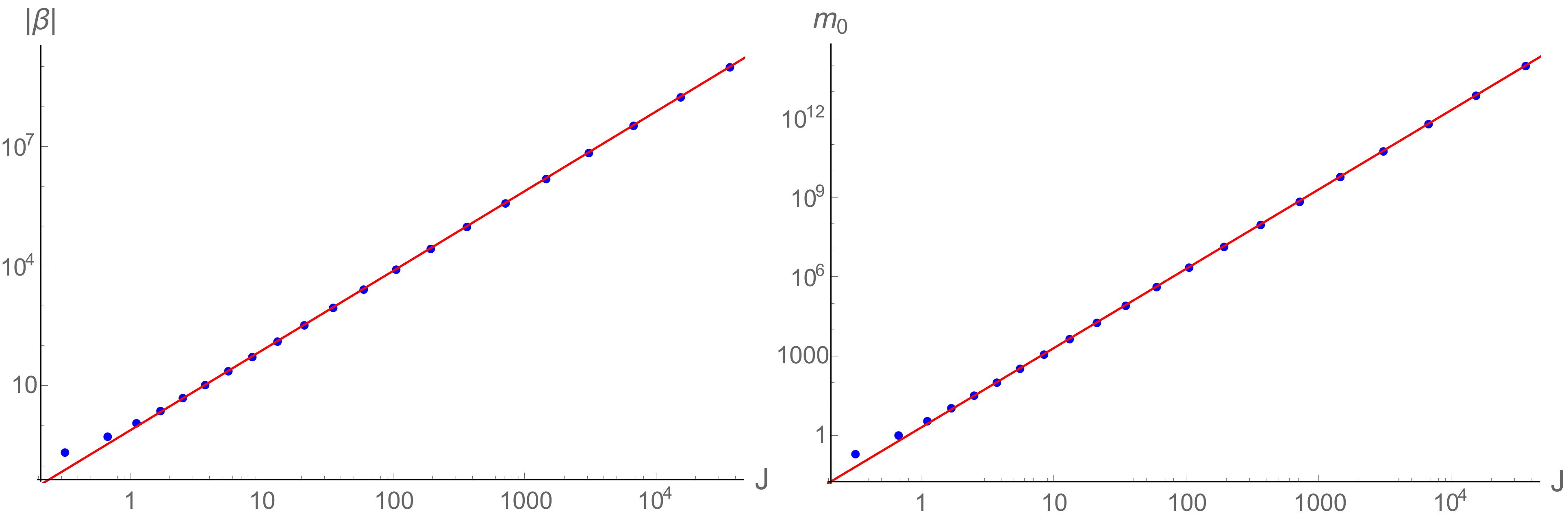}
		\caption{\small Left: $|\epsilon(J)|$ together with its fit to a $J^2$ power law. Right: $m_0(J)$ together with its fit to a $J^3$ power law.}
		\label{fig:planar_soliton}
	\end{center}
\end{figure}

In order to exemplify this procedure, let us consider the soliton family. In Fig.\,\ref{fig:planar_soliton}, we plot $|\epsilon(J)|$ and $m_0(J)$ along the soliton branch, together with the fit of the last ten points to 
\beq
\epsilon(J) = \hat \epsilon J^2,~~~~~~~m_0(J) = \hat m_0 J^3. 
\eeq
The agreement between the behavior of these quantities and the expectations drawn from the existence of a nontrivial planar limit is excellent. In particular, from the fits we obtain that\footnote{As an aside, note that the First Law \eqref{first_law} implies that ($s = 0$ for any soliton solution)
	\beq
	1/2 dm_0 + 2 \epsilon dJ + J d\epsilon = (3/2 \hat m_0 + 4 \hat \epsilon)J^2 dJ = 0. 
	\eeq
	By knowing $\hat \epsilon$, we automatically know $\hat m_0 = - 8/3 \hat \epsilon$. Indeed, $1.9734 \approx 8/3 \times 0.7400$.}
\beq
\hat \epsilon \approx -0.7400,~~~~~~~\hat m_0 \approx 1.9734. 
\eeq

These observations establish without a trace of doubt that the soliton branch has a nontrivial planar limit. What remains to be done now is to construct the planar, limiting solution directly, and not through a blow-up limit. This is the objective of the next subsection.

\subsubsection{The planar limit of the AdSJ branch}

We are working with planar asymptotics now. Our first working hypothesis  is that the planar geometry to which the soliton branch maps upon blow-up is Poincar\'e invariant. We also know that this putative solution is static and supported by a nontrivial scalar field profile, as its soliton progenitors. As a second working hypothesis, we are going to assume that this scalar field profile is monotonic, i.e., that 
\beq
\phi'(\chi) > 0, 
\eeq
where $\chi$ is a suitable radial coordinate. This is natural to the extent that it also holds for the each individual solution in the soliton branch. Monotonicity implies that we can parameterize the radial direction in terms of $\phi$ itself, i.e., the implicit function theorem allows us to write $\chi = \chi(\phi)$. 

These considerations motivate us to choose the metric ansatz
\beq
ds^2 = e^{A(\phi)}(-dt^2 + dx^2 + dy^2) + B(\phi) d\phi^2. \label{planar_ansatz}
\eeq
From the Einstein equation, we get that 
\beqa
&&B(\phi) = \frac{3 A'(\phi)^2 - 4}{4(3+2\phi^2)}, \label{Beq} \\
&&A''(\phi) = \frac{(3+2\phi^2+\phi A'(\phi))(-4+3 A'(\phi))}{6+4 \phi^2}, \label{Aeq}
\eeqa
while these relations automatically solve the scalar field equation. 

In order to solve the equation of motion of $A$, we need to impose suitable boundary conditions. We know that the asymptotic UV region corresponds to the limit $\phi \to 0$, but what do the IR region corresponds to? 

Regarding this point, note first that the planar solution must have zero entropy density, i.e., the area element of a constant time, constant $\phi$ hypesurface must tend to zero as such hypersurface tends to the far IR of the geometry. This implies that, as $\phi \to \phi_{IR}$, $A(\phi)$ must tend to $-\infty$. Here $\phi_{IR}$ is the value of the scalar field in the far IR of the geometry; at present we don't wether $\phi_{IR}$ is finite or $\infty$.\footnote{We are not considering the case that $\phi_{IR} = 0$ since, as $\phi$ must also vanish at the boundary, this implies that $\phi$ is not monotonic and violates one of our assumptions.} 
\\\\
\noindent In the case that $\phi_{IR}=\infty$, is easy to find a solution. If we assume that $A(\phi) \sim - \phi^2$ as $\phi \to \infty$, eq. \eqref{Aeq} can be solved in a series expansion in inverse powers of $\phi$, together with an extra logarithmic  contribution. Explicitly,
\beq
A(\phi) = - \phi^2 - \frac{11}{3} \log \phi - \frac{11}{9 \phi^2} + \frac{55}{27 \phi^4} - \frac{671}{108 \phi^6} + O\left(\frac{1}{\phi^8} \right). 
\eeq
Then, by eq. \eqref{Beq}, 
\beq
B(\phi) = \frac{3}{2} + \frac{11}{4 \phi^2} - \frac{11}{4 \phi^4}+ \frac{77}{8 \phi^6} + O\left(\frac{1}{\phi^8} \right).  
\eeq
In this case, $B$ is finite in the IR. From these series expansions, we clearly see that the IR solution is uniquely fixed and has no adjustable parameters.\footnote{It should be noted that $A$ only appears in its equation of motion through its derivatives and, as a consequence, there exists the residual freedom of shifting it by a constant $a_0$. This shift can be undone by a coordinate change that modifies $t, x, y$ by a multiplicative factor and has no invariant meaning. In the expression of the main text we have set this constant to zero.} It should also be noted that this solution is singular in the IR. For instance, the Ricci scalar behaves as 
\beq
R = - 8 \phi^2 - \frac{32}{3} + O\left(\frac{1}{\phi^2} \right),   
\eeq
while the Kretschmann scalar as
\beq
K = \frac{32}{3} \phi^4 + \frac{256}{9} \phi^2 + \frac{760}{27} + O\left(\frac{1}{\phi^2} \right). 
\eeq
Is this singularity naked? To determine this, we must find out if there exists an apparent horizon in the solution.\footnote{Since the metric is time-independent, this horizon must correspond to the event horizon.} However, the series expansions are apparently not sufficient at this point: it seems that we need to find the full metric. This is not the case. We will show this by understanding better the IR geometry. In order to so, it is illustrative to change our coordinate system. From \eqref{planar_ansatz}, we see that the Schwarzschild radial coordinate is given by \beq
r(\phi) = e^{1/2 A(\phi)}. \label{r_phi_2}
\eeq
Note that, in particular, $r \to 0$ as $A \to -\infty$. To lowest nontrivial order in the $\phi \to \infty$ limit, 
\beq
r(\phi) \sim \frac{e^{-\frac{1}{2}\phi^2}}{\phi^\frac{11}{6}}, \label{r_phi}
\eeq
from where we can solve for $\phi(r)$ in terms of the Lambert W function, 
\beq
\phi(r) \sim \sqrt{\frac{11}{6}} \sqrt{W\left(\frac{6}{11 r^\frac{12}{11}}  \right)}. 
\eeq
This expression is only valid at leading order as $r  \to 0$. In this limit, the argument of the W function diverges. Therefore, we can employ the leading order asymptotic expansion of W, 
\beq
W(x) = \log x + O(\log \log x), 
\eeq
to get\footnote{This result also shows that neglecting the power law term in \eqref{r_phi} is justified, since we could have obtained the same result by not including it.}
\beq
\phi(r) \sim \sqrt{2} \sqrt{- \log r}. 
\eeq
Although $\phi$ diverges as $r \to 0$, this divergence is rather mild. In terms of the $r$ coordinate, $B(\phi)$ behaves as 
\beq
B(r) = B(\phi(r)) \sim 3/2
\eeq
Therefore, at leading order, the $rr$-component of the metric tensor is 
\beq
g_{rr} = B(r) \phi'(r)^2 \sim \frac{3}{4} \frac{1}{r^2 (-\log r)}
\eeq
and the whole metric asymptotes to the following IR geometry, 
\beq
ds^2 \sim r^2 (-dt^2 + d{\vec x}^2) + \frac{dr^2}{\frac{4}{3} r^2 (- \log r)}. 
\eeq
The scale invariance of the Poincar\'e patch is mildly broken by the $\log r$ term: only Poincar\'e invariance remains. Metrics of this kind have been previously found to be zero-temperature limits of holographic superconductors [HR]. Comparing this expression with the standard metric ansatz,
\beq
ds^2 = -f(r) e^{-2 \delta(r)}dt^2 + \frac{dr^2}{f(r)} + r^2 d\vec x^2
\eeq
we see that 
\beqa
f(r) \sim -\frac{4}{3}r^2 \log r, \\
\delta (r) \sim \frac{1}{2} \log\left(-\frac{4}{3}\log r \right). 
\eeqa
In particular, $f(r=0) = 0$, so the $r=0$ hypersurface corresponds to an apparent horizon. Therefore, the singularity is naked. It is also possible to compute the temperature associated to this horizon by the standard relation\footnote{Usually there is no contribution from the $\delta'$ term since it is finite and comes multiplied by $f$, which is vanishing at the horizon. This is not the case here since $\delta'$ diverges.} 
\beq
T = \frac{1}{4 \pi} \sqrt{f'(r) \left(f(r) e^{-2 \delta(r)}\right)'}\bigg\rvert_{r=0}. 
\eeq
We get that, as $r \to 0$, 
\beq
T \sim \frac{r \sqrt{- 2 \log r}}{\sqrt{6}\pi}, 
\eeq
and so $T = 0$. This solution has strict zero temperature. If the singular solution corresponds to the planar limit of the soliton branch, the Hawking-Page transition cannot have a planar limit. The reason is that the phase which dominates the grandcanonical ensemble in global AdS at low but finite temperatures maps to a zero temperature solution; in planar AdS, at any fixed and finite temperature, there is only one phase: the one that corresponds to the hairy black hole solutions. There are no competing saddles and no phase transition. This observation leads to the prediction that, when $J \to \infty$, $T_{HP}/J$ must vanish.
\\\\
\noindent The previous discussion shows how important it is to find out whether the singular solution is the blow-up limit of the soliton branch. To do this, we solve \eqref{Aeq} numerically by shooting from $\phi_{IR} \gg 1$ to $\phi_{UV} \ll 1$, using the IR series expansion of $A$ to provide the initial conditions for the shooting. Once $A(\phi)$ is known, $\phi(r), f(r)$ and $\delta(r)$ can be easily determined. Let us explain how. 

\begin{figure}[h!]
	\begin{center}
		\includegraphics[width=16cm]{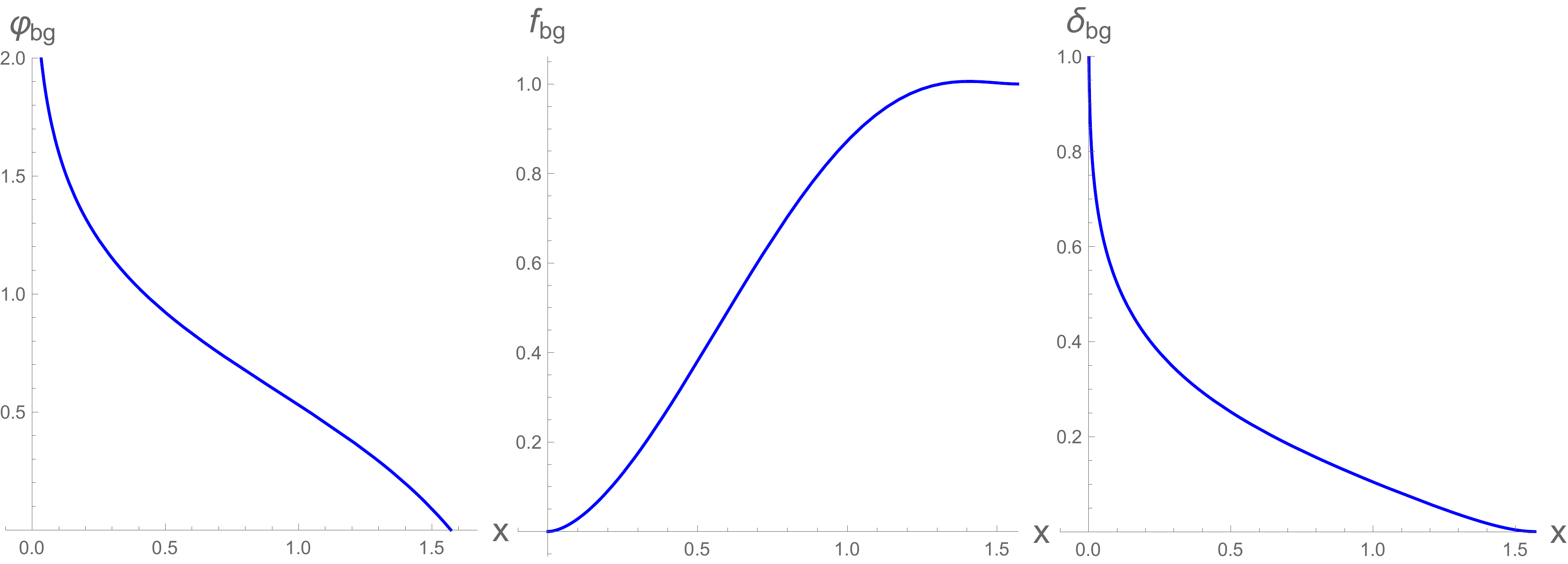}
		\caption{\small Metric and scalar field of the zero temperature solution}
		\label{fig:planar_bg}
	\end{center}
\end{figure}

Let us compactify the radial direction as usual,  $r = \tan x$. Note that $x(\phi) = \arctan e^{1/2 A(\phi)}$. In terms of $\phi$, we have that 
\beqa
&&f(\phi) = \sec^2(x(\phi)) \tilde f(\phi), \tilde f(\phi) = \frac{e^{A(\phi)}(3+2\phi^2)A'(\phi)^2}{(1+e^{A(\phi)})(3 A'(\phi)^2-4)} \\
&&\delta(\phi) = \frac{1}{2} \log \frac{(3+2\phi^2)A'(\phi)^2}{3 A'(\phi)^2 - 4}. 
\eeqa
where $\tilde f$ is analogous to the emblackening factor we have been using in the global case. Once these expressions are known, we interpolate $(x(\phi), \phi)$,  $(x(\phi), \tilde f(\phi))$ and  $(x(\phi), \delta(\phi))$. The results we have obtained are plotted in Fig.\, \ref{fig:planar_bg}. 

Once the background is known, we can compute the dimensionless ratios \eqref{ratios}, with the result
\beq
\frac{\epsilon}{J^2} = 0.7394,~~~~~~\frac{m_0}{J^3} = 1.9741. 
\eeq
These ratios are extremely close to the values we have determined independently by analyzing the large $J$ behavior of $\epsilon$ and $m_0$ along the soliton branch. In particular, 
the relative difference between both $\hat \epsilon$'s is 0.081$\%$ , while the relative difference between both $\hat m_0$'s is 0.035$\%$: both quantities have a relative deviation smaller than $0.1\%$. 

This establishes without a trace of doubt that the zero temperature planar solution we have constructed is the planar limit of the soliton branch, being the very small differences we observe due to accumulated numerical error.

According to our previous discussion, the fact that the planar limiting solution has strictly zero temperature implies that the Hawking-Page transition does not have a planar limit in this model. In particular, we must have that $\lim_{J\to \infty} T_{HP}/J = 0$. 

\bibliographystyle{unsrt}
\bibliography{draft1}
\end{document}